\documentclass{aa} 

\usepackage[varg]{txfonts}
\usepackage{graphicx}

\usepackage[dvipsnames]{xcolor}
\usepackage[colorlinks,citecolor=RoyalBlue,linkcolor=blue]{hyperref}
\usepackage{placeins}
\usepackage{etoolbox}
\begin{document}

\title{Deciphering stellar metallicities in the early Universe: Case study of a young galaxy at $z=4.77$ in the MUSE eXtremely Deep Field\thanks{Based on observations obtained with the Very Large Telescope under the large program 1101.A-0127.}}
\titlerunning{Metallicity of a young starburst at $z=4.77$ in the MXDF}
\authorrunning{Matthee et al.}

\author{Jorryt Matthee\inst{1}\thanks{Zwicky Fellow. E-mail: mattheej@phys.ethz.ch},
Anna Feltre\inst{2}, 
Michael Maseda\inst{3,4},
Themiya Nanayakkara\inst{5},
Leindert Boogaard\inst{6,3},
Roland Bacon\inst{7},
Anne Verhamme\inst{8},
Floriane Leclercq\inst{8},
Haruka Kusakabe\inst{8},
Tanya Urrutia\inst{9} 
 \and Lutz Wisotzki\inst{9}}

%\offprints{ZZZ, \email{mailaddress}

\institute{Department of Physics, ETH Z\"urich, Wolfgang-Pauli-Strasse 27, 8093 Z\"urich, Switzerland
  \and INAF - Osservatorio di Astrofisica e Scienza dello Spazio di Bologna, Via P. Gobetti 93/3, 40129, Bologna, Italy 
  \and Leiden Observatory, Leiden University, PO\ Box 9513, NL-2300 RA Leiden, The Netherlands
  \and Department of Astronomy, University of Wisconsin-Madison, 475 N. Charter St., Madison, WI 53706 USA
  \and Centre for Astrophysics and Supercomputing, Swinburne University of Technology, Hawthorn, Victoria 3122, Australia.
  \and Max Planck Institute for Astronomy, K\"{o}nigstuhl 17, 69117, Heidelberg, Germany 
     \and Univ. Lyon, Univ. Lyon1, ENS de Lyon, CNRS, Centre de Recherche Astrophysique de Lyon UMR5574, 69230 Saint-Genis-Laval, France
  \and Observatoire de Gen\`eve, Universit\'e de Gen\`eve, Chemin Pegasi 51, 1290 Versoix, Switzerland          
  \and  Leibniz-Institut f\"{u}r Astrophysik Potsdam (AIP), An der Sternwarte 16, 14482 Potsdam, Germany   } 

%\date{Received ZZZ / Accepted ZZZ}

\abstract{
Directly characterising the first generations of stars in distant galaxies is a key quest of observational cosmology. We present a case study of ID53 at $z=4.77$, the UV-brightest (but L$^{\star}$) star-forming galaxy at $z>3$ in the MUSE eXtremely Deep Field with a mass of $\approx10^9$ M$_{\odot}$. In addition to very strong Lyman-$\alpha$ (Ly$\alpha$) emission, we clearly detect the (stellar) continuum and an N{\sc v} P-Cygni feature, interstellar absorption, fine-structure emission and nebular C{\sc iv} emission lines in the 140 hr spectrum. Continuum emission from two spatially resolved components in Hubble Space Telescope data are blended in the MUSE data, but we show that the nebular C{\sc iv} emission originates from a subcomponent of the galaxy. The UV spectrum can be fit with recent BPASS stellar population models combined with single-burst or continuous star formation histories (SFHs), a standard initial mass function, and an attenuation law. Models with a young age and low metallicity (log$_{10}$(age/yr)=6.5-7.6 and [Z/H]=-2.15 to -1.15) are preferred, but the details depend on the assumed SFH. The intrinsic H$\alpha$ luminosity of the best-fit models is an order of magnitude higher than the H$\alpha$ luminosity inferred from {\it Spitzer}/IRAC data, which either suggests a high escape fraction of ionising photons, a high relative attenuation of nebular to stellar dust, or a complex SFH. The metallicity appears lower than the metallicity in more massive galaxies at $z=3-5$, consistent with the scenario according to which younger galaxies have lower metallicities. This chemical immaturity likely facilitates Ly$\alpha$ escape, explaining why the Ly$\alpha$ equivalent width is anti-correlated with stellar metallicity. Finally, we stress that uncertainties in SFHs impose a challenge for future inferences of the stellar metallicity of young galaxies. This highlights the need for joint (spatially resolved) analyses of stellar spectra and photo-ionisation models.}

\keywords{Galaxies: high-redshift -- Techniques: spectroscopic -- Galaxies: stellar content -- Galaxies: formation} 
\maketitle

\section{Introduction} \label{sec:intro}
Identifying galaxies that host the first generations of stars is a major goal of extragalactic astrophysics. This is likely an extremely difficult challenge because very faint low-mass galaxies in the distant Universe need to be observed in order to see galaxies whose light is dominated by these first-generation stars with metallicities $Z\lesssim10^{-4} Z_{\odot}$ \citep[e.g.][]{Schauer2020}. The redshift out to which such systems exist depends on poorly understood properties such as the efficiency of metal mixing and radiative feedback of nearby haloes \citep{Scannapieco2003,Xu2016,Visbal2017,Liu2020}. As a single pair-instability supernova already enriches the surrounding gas to a metallicity of $Z\sim10^{-3} Z_{\odot}$ \citep{Wise2012}, it is unlikely that we will be able observe purely metal-free galaxies. It is thus plausible that systems with mixed populations of primordial and more metal-rich Population II stars exist \citep[e.g.][]{Tornatore2007,Pallottini2015}. Observational constraints of galaxies with low stellar metallicities in galaxies can chart these enrichment and mixing processes. 

The evolution of the stellar metallicity in galaxies is also of interest as  a tracer of outflows and gas recycling \citep[e.g.][]{Weinberg2017}. It is important to know the stellar metallicity of distant galaxies in order to observationally infer other properties accurately. The hardness of stellar atmospheres depends on their metallicity (i.e. metal-poor stars are hotter). This impacts the strengths of various nebular emission lines \citep[e.g.][]{Steidel2016,Topping2020}, which are important to infer the gas-phase metallicity \citep[e.g.][]{Sanders2020} and even the star formation rate \citep[e.g.][]{Theios2019}. Furthermore, understanding the stellar metallicity and its cosmic evolution is of key importance for understanding the hosts of gravitational wave events \citep[e.g.][]{Chruslinska2019,LIGO2020}.

Recent searches for galaxies that contain very metal-poor stars have targeted galaxies with very blue colours \citep[e.g.][]{Bouwens2010,Bhatawdekar2020}. This is an efficient method as it only requires multi-band photometry, but it is not very specific. Moreover, extremely young stellar populations may even have relatively red colours due to the contribution from nebular continuum \citep[e.g.][]{Raiter2010}. Strong nebular emission from the hydrogen Lyman-$\alpha$ (Ly$\alpha_{1216}$) and, in particular, the helium Balmer-$\alpha$ (He{\sc ii}$_{1640}$) recombination lines has also been proposed as a tracer of very massive metal-poor stars \citep{Tumlinson2001,Schaerer2002}, inspiring early attempts of identifying these lines in relatively massive galaxies over $z=3-7$ \citep{Cai2011,Cassata2013,Cai2015,Sobral2015,Sobral2019}. 

Nebular  He{\sc ii}$_{1640}$ emission, however, can also be powered by stars with low, but not primordial, metallicities \citep[e.g.][]{GrafenerVink2015,Szecsi2015}, or non-thermal sources such as X-ray binaries, radiative shocks, or active galactic nucleii (e.g. \citealt{Plat2019,Schaerer2019}; c.f. \citealt{Senchyna2020}). Narrow He{\sc ii} line emission has been observed in a range of individual and stacked high-redshift galaxies that are clearly not primordial \citep[e.g.][]{Fosbury2003,Erb2010,Steidel2016,Berg2018,Nanayakkara2019,Saxena2020,Matthee2021} and in their local analogues \citep[e.g.][]{Kehrig2015,Berg2016,Berg2019,Senchyna2017,Wofford2021}. In practically all cases, these He{\sc ii} lines appear together with strong emission from high-ionisation metal lines such as C{\sc iv} and O{\sc iii}] \citep[e.g.][]{Stark2014,Nanayakkara2019}. As \cite{StanwayEldrige2019} showed, the relative strength of the observed He{\sc ii} lines is challenging if not impossible to explain with the most recent stellar models, suggesting that an additional process contributes to the ionisation. 

The most direct way to observationally constrain the properties of massive stars in galaxies is through sensitive spectroscopy of the starlight. Because of sensitivity requirements, such studies are typically limited to observations at low redshift \citep[e.g.][]{Chisholm2019}, stacking techniques \citep{Steidel2016,Cullen2019,Calabro2021}, or observations of highly magnified galaxies \citep{Patricio2016,Chisholm2019}. From rest-frame UV observations, the stellar metallicity can be inferred through its impact on photospheric line-blanketing \citep{Rix2004,Sommariva2012} and broad and P-Cygni features of high-ionisation lines such as C{\sc iv}, He{\sc ii}, O{\sc v,} and N{\sc v} \citep[e.g.][]{Brinchmann2008,Byler2018,Chisholm2019}. The main challenges for this approach are uncertainties in the initial mass function (IMF), the stellar models, and the star formation history (SFH). Flexible star formation histories are particularly important in the case of clumpy galaxies that are frequently observed at high redshift \citep[e.g.][]{Matthee2017ALMA,Matthee2020,Carniani2018,Herrera-Camus2021}. 

In this paper we present a case study of how well we can infer the stellar age and particularly the metallicity from a high-quality rest-frame UV spectrum of an individual unlensed galaxy in the early Universe. We study the galaxy ID53 at $z=4.77$ detected in the MUSE Deep Field \citep{Bacon2017} observed with the Multi Unit Spectroscopic Explorer (MUSE; \citealt{Bacon2010}). Its Ly$\alpha$ emission has previously been studied by \cite{Leclercq2017}. We use extremely deep spectroscopic data from the VLT/MUSE eXtremely Deep Field (MXDF) with an on-source exposure time of 140 hours. While ID53 is a typical L$^{\star}$ galaxy, we study it here because it is the brightest observed galaxy at $z>3$ in the MXDF data. This case study explores the practical uncertainties and issues in deriving stellar metallicities from sensitive rest-frame UV spectroscopy (but limited from $\lambda_0=1220-1600$ {\AA} by the wavelength coverage of MUSE and the redshift of ID53) combined with (near)-infrared photometry, in order to guide future observing strategies, for example those with the Extremely Large Telescopes. 

In \S $\ref{sec:data}$ we present the data and relevant measurements. We introduce the general properties of the galaxy in  \S $\ref{sec:what}$, including the UV luminosity and slope and the rest-frame UV morphology from existing data. Here we also present the measurement of the systemic redshift (from non-resonant fine-structure lines) and the detection of nebular C{\sc iv} emission. We then focus on fitting the stellar continuum spectrum using BPASS v2.2 \citep{BPASS2018} models in \S $\ref{sec:stellar_model}$ with an emphasis on the uncertain SFH. In \S $\ref{sec:nebular}$ we discuss how measurements of the H$\alpha$ luminosity may help constrain different models, and we present an inference of the H$\alpha$ luminosity of ID53 through {\it Spitzer}/IRAC data. We compile the various observational clues on ID53 into a physical picture of the ongoing starburst in \S $\ref{sec:synthesis}$. In \S $\ref{discuss:EW}$ we then combine our results with those in the literature to show that galaxies with strong Ly$\alpha$ emission are good targets to follow up in order to find galaxies that are dominated by very metal-poor star formation. In \S $\ref{sec:discuss}$ we discuss how further future observations are useful in overcoming the current limitations and caveats. Finally, we summarise our results in \S $\ref{sec:summary}$.  

Throughout this work, distances and luminosities are calculated assuming a flat $\Lambda$CDM cosmology with $\Omega_{\Lambda, 0}=0.7$, $\Omega_{\rm M, 0}=0.3,$ and H$_{0}=70$ km s$^{-1}$ Mpc$^{-1}$. Magnitudes are in the AB system \citep{OkeGunn1983}. Stellar metallicities are expressed relative to the solar abundance as [Z/H]=log$_{10}$($Z/Z_{\odot}$), where we use $Z_{\odot}=0.014$ \citep{Asplund2009}.

\begin{table}
\centering
\caption{General properties of ID53.} \label{tab:photometry}
\begin{tabular}{lc} \hline
R.A. & 3:32:37.952 (J2000) \\
Dec. & -27:47:10.93 (J2000) \\
$z_{\rm sys}$ & $4.7745\pm0.0004$ \\ 
M$_{1500}$ &  $-21.07$ \\
$\beta$ & $-2.01\pm0.05$ \\
L$_{\rm Ly\alpha}$ & $1.48\times10^{43}$ erg s$^{-1}$ \\
EW$_{\rm 0, Ly\alpha}$ & $62\pm3$ {\AA} \\
\hline
F105W & $25.30\pm0.03$ \\
F125W & $25.25\pm0.03$ \\
F160W & $25.28\pm0.03$ \\
$K_s$ & $25.24\pm0.10$ \\
{[3.6]} & $24.82\pm0.05$\\
{[4.5]} & $25.47\pm0.09$ \\
\hline
\end{tabular}
\end{table}

\begin{figure*}
        \includegraphics[width=18.6cm]{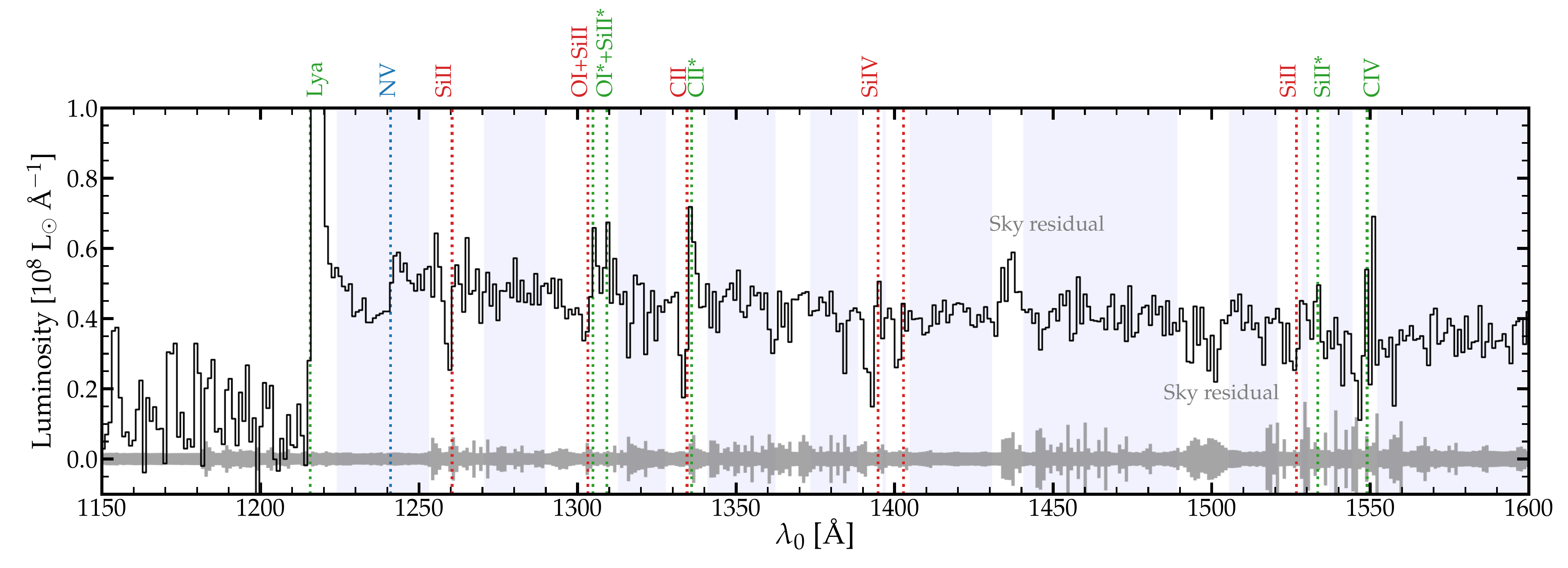} 
    \caption{Rest-frame 1D spectrum of ID53 extracted from the MXDF data (black line), binned by a factor 5 in the wavelength direction. The spectrum is shifted to the rest-frame based on $z_{\rm sys}=4.7745$ and scaled to rest-frame luminosity density. The range of the y-axis is limited for visualisation purposes. The grey shaded region shows the noise level and particularly visualises the locations of skylines. Dotted red lines mark the expected locations of interstellar absorption lines, dotted green lines mark emission lines, and the dotted blue line marks the stellar N{\sc v} feature. We also highlight two wavelength regions that are affected by significant skyline residuals. The blue shaded regions highlight spectral regions that have been included in the fitting of the stellar continuum.}
    \label{fig:1dspec}
\end{figure*}
  
\section{MUSE data} \label{sec:data}
The spectrum of ID53 that we analyse in this paper was extracted from data in the MUSE eXtremely Deep Field (see \citealt{Bacon2021} for a detailed description). The MXDF is a recently completed campaign that observed a circular region with a radius of 31$''$  in the {\it Hubble} eXtreme Deep Field \citep{Illingworth2013} with an exposure time of $100+$ hours (140 hours on the location of ID53) with MUSE on the VLT. The observations and data reduction are presented in Bacon et al. (in prep). The observations were assisted by ground-layer adaptive optics and therefore have an excellent image quality with a point spread function that has a relatively constant full width at half maximum (FWHM) $\approx0.5''$ (with a Moffat profile with $\beta=1.96$) over $\lambda=700-930$ nm, which is the wavelength region of interest in this work. MUSE data-cubes consist of 3721 wavelength-layers spanning $\lambda=470-935$ nm with a width of 1.25 {\AA.}  The spectral resolution FWHM over the wavelength range of interest for this work is $R\approx3000$ \citep{Bacon2017}. 

We extracted the spectrum of ID53 based on the spatial profile of the UV continuum emission measured with MUSE, which is marginally resolved in the data. We collapsed the data-cube over $\lambda_{\rm obs}=715-865$ nm ($\langle \lambda \rangle=790$ nm), which corresponds to the rest-frame wavelengths $\lambda_0=124-150$ nm that are not contaminated by strong line-emission, and fit a 2D Gaussian profile to the image. The best-fit profile has an FWHM=0.67$''$, 0.84$''$ along the minor and major axes, respectively, with a position angle of 48 degrees. The 1D spectrum was extracted by using the Gaussian profile as a weight map. The FWHM of the weight map is wavelength-dependent following the (weak) wavelength dependence of the PSF FWHM compared to the reference wavelength of $\langle \lambda \rangle=790$ nm. The noise level of the 1D spectrum was propagated from the variance cube associated with the data. 

The 1D spectrum of ID53 is shown in Fig. $\ref{fig:1dspec}$, where we binned the spectrum by a factor 5 in the wavelength direction for visualisation purposes. Without binning, the UV continuum is detected with a signal-to-noise ratio (S/N) ranging from 6-15 per layer of $\Delta\lambda_{\rm obs}=1.25$ {\AA} that is not contaminated by skyline emission. For most of the analyses of the paper, we binned the spectrum by a factor 5 because this corresponds to a rest-frame slicing of $\Delta\lambda_{0}\approx1$ {\AA} ($\approx220$ km s$^{-1}$), which is similar to the sampling of the stellar population models considered in our analysis. In this case, the S/N of the continuum is a factor $\approx\sqrt{5}$ higher, with a slightly lower increase in wavelength regions near skylines.

As the Lyman-$\alpha$ emission is spatially significantly more extended than the UV continuum emission, our spectral extraction underestimates the Lyman-$\alpha$ flux. Therefore, we used the Ly$\alpha$ flux and equivalent width measurements from \cite{Leclercq2017} that are based on a spatially resolved analysis of data with a total integration time of 30 hours (see Table $\ref{tab:photometry}$). \cite{Leclercq2017} separated the contributions from the Ly$\alpha$ core (i.e. Ly$\alpha$ luminosity with a similar spatial distribution as the UV continuum) and the extended diffuse Ly$\alpha$ halo luminosity. They measured that 45 \% of the Ly$\alpha$ luminosity is observed as halo flux. The Ly$\alpha$ equivalent width (EW) in our 1D spectrum (extracted over the PSF-convolved UV continuum profile) is $55\pm1$ {\AA}. This is only slightly lower than the total Ly$\alpha$ EW of 62 {\AA} that takes diffuse extended Ly$\alpha$ emission into account.

\section{Properties of the galaxy ID53} \label{sec:what}

\subsection{General properties} \label{sec:phot}
\begin{figure}
\centering
        \includegraphics[width=9.3cm]{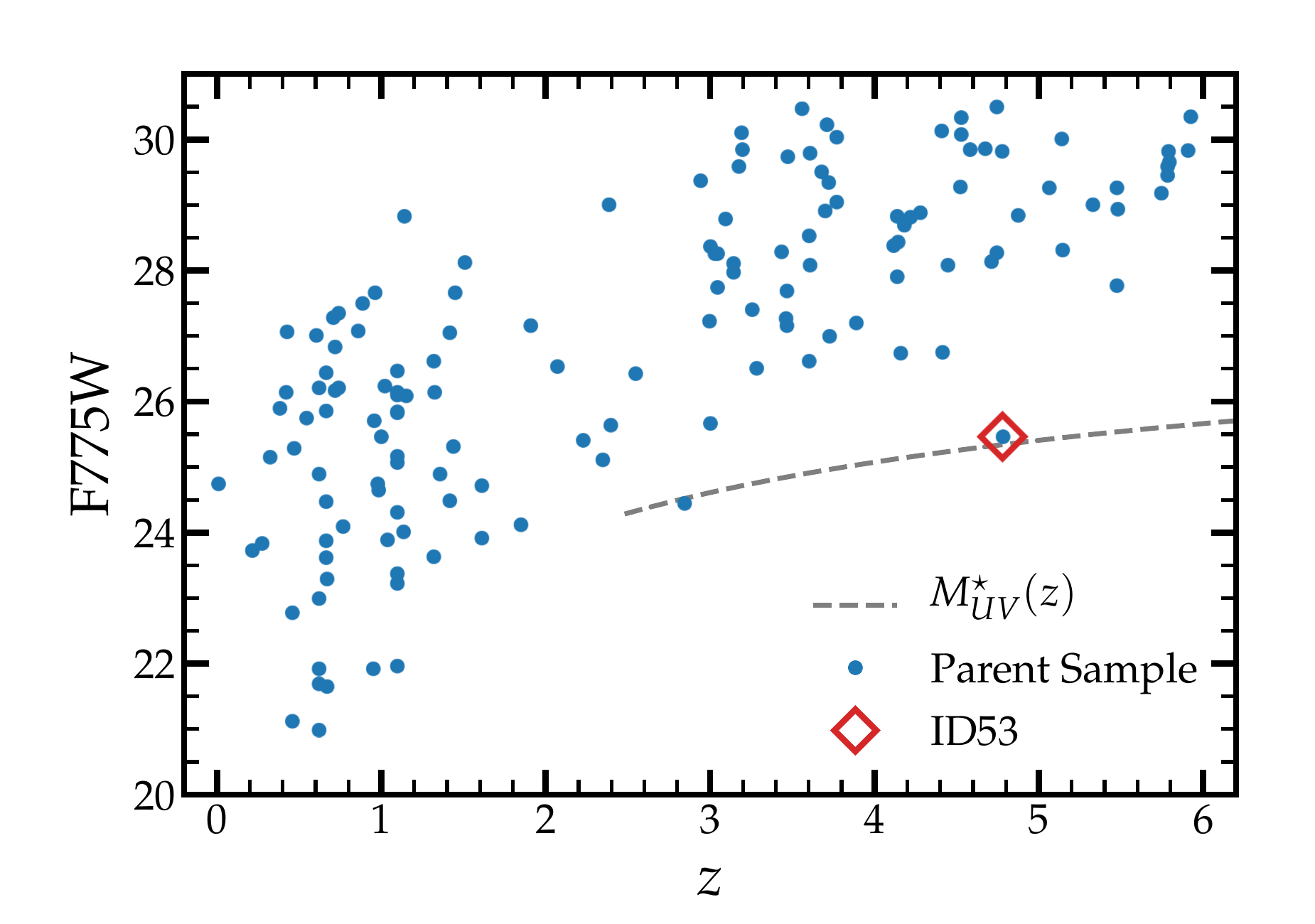} 
    \caption{Observed {\it HST}/ACS F775W magnitudes vs. the spectroscopic redshifts of galaxies within the region of the MXDF with a total exposure time of $>100$ hours. The dashed grey line shows the evolution of the characteristic luminosity of the UV luminosity function \citep{Bouwens2021}. ID53 stands out by its continuum luminosity, which is the brightest at $z>3$ and 1.5 magnitudes, brighter than any other galaxy at $z\approx5$ in these data. }
    \label{fig:sample}
\end{figure}

The galaxy ID53 is observed at $z=4.77,$ and it stands out by its continuum luminosity compared to other galaxies in the MXDF field. As shown in Fig. $\ref{fig:sample}$, ID53 is the brightest observed galaxy in the data at $z>3,$ and it is more than 1.5 magnitudes brighter than other galaxies at comparable redshift. Despite this relatively high luminosity, its UV luminosity is comparable to L$^{\star}_{\rm UV}$ \citep{Bouwens2015}. ID53 also has a bright Lyman-$\alpha$ (Ly$\alpha$) line with a luminosity comparable to L$^{\star}_{\rm Ly\alpha}$ \citep{SC4K,Herenz2019} and a relatively high Ly$\alpha$ EW$_0 = 62$ {\AA}. The UV slope is $\beta=-2.01$ (measured with the F105W, F125W, and F140W filters; \citealt{Hashimoto2017}), which is relatively blue given its luminosity (typical $\beta=-1.7$ for this UV luminosity at $z\approx5$; \citealt{Bouwens2014slope}). The stellar mass of ID53 is $\approx3\times10^9$ M$_{\odot}$ (\citealt{Santini2015}, but see also our fitting in \S $\ref{sec:stellar_model}$). For a \cite{Chabrier2003} IMF, the star formation rate associated with the unobscured UV luminosity is 10 M$_{\odot}$ yr$^{-1}$ \citep{Kennicutt1998}. This implies that the star formation rate (SFR) of ID53 is typical for its stellar mass at $z\approx5$ \citep{Speagle2014} as long as the unobscured SFR is at least one-third of the total SFR. This is plausible given the blue UV slope \citep[e.g.][]{Meurer1999}. Our fitting suggests an E($B-V$)$\approx0.2,$ which places the galaxy in the upper envelope of the SFR-M$_{\rm star}$ relation.

\begin{figure}
\centering
        \includegraphics[width=8.3cm]{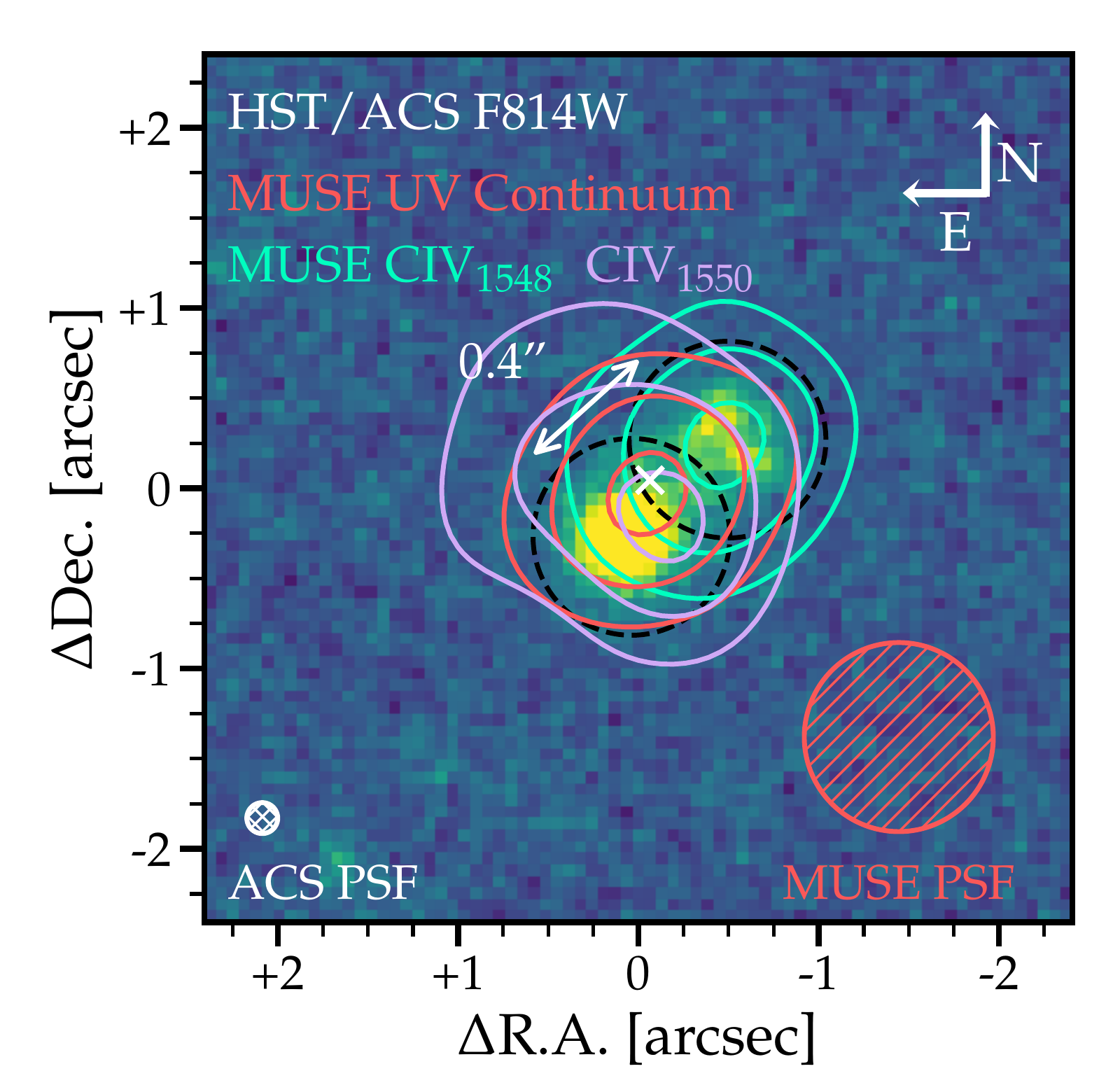} 
    \caption{{\it HST}/ACS F814W image of ID53 from the eXtreme Deep Field project (\citealt{Illingworth2013}). The F814W filter traces the rest-frame UV emission that is also traced by the MUSE spectrum. The PSF-FWHM of the MUSE data is shown in the bottom right corner for comparison. The white cross marks the centre of the Ly$\alpha$ emission. ID53 consists of (at least) two components in the high-resolution imaging. These are significantly blended in the MUSE continuum (red). Spatial offsets are seen for pseudo-narrow bands of the C{\sc iv}$_{1548,1550}$ emission. The bluer line (shown in cyan) spatially peaks at the fainter UV component, but the C{\sc iv}$_{1550}$ line peaks at the position of the continuum, which is suggestive of strongly overlapping C{\sc iv} lines at slightly different redshifts. Dashed black circles show the FWHM of the extraction apertures of point sources in the NW and SE in the MUSE data, respectively. These extractions are shown in Fig. $\ref{fig:carbon_resolved}$.} 
    \label{fig:HST}
\end{figure}

In addition to the MUSE data, ID53 is covered by extremely deep {\it HST}, {\it Spitzer,} and ground-based imaging data. For the purpose of this paper, we used photometry based on these data to help constrain the SED models, in particular in redder wavelengths than those covered by the MUSE data. Some of these data (marginally) resolve ID53 (see Fig. $\ref{fig:HST}$), but we only used the integrated photometry of the system. We collected total flux measurements based on near-infrared data in the F105W, F125W, and F160W filters from {\it HST}/WFC3 by \cite{Rafelski2015}, $K_s$ -band data from FORS2 on the VLT from \cite{Fontana2014}, and mid-infrared data in the [3.6] and [4.5] filters on {\it Spitzer}/IRAC from the GREATS program \citep[e.g.][]{Labbe2015,DeBarros2019}. The measurements are listed in Table $\ref{tab:photometry}$. We note that we added a 0.03 magnitude uncertainty in quadrature to these measurements to conservatively account for uncertainties in the absolute flux calibration and aperture corrections. ID53 is not detected in any other band or wavelength (e.g. X-Ray, radio, sub-millimeter) besides those listed in Table $\ref{tab:photometry}$.

\subsection{High-resolution morphology} \label{sec:morph}
In the high-resolution {\it HST}/ACS images (FWHM$\approx0.08''$, about eight times better than the MUSE resolution), ID53 is resolved in at least two, potentially three, components, see Fig. $\ref{fig:HST}$. The south-eastern (SE) and north-western (NW) components are separated by 0.4$''$ (2.6 kpc) and have relative flux ratios of $\approx1:2$ in the UV continuum traced by the F814W filter shown here. The south-eastern component is the brightest. The north-western component appears to be clumpy in itself as well. We note that spatially varying dust attenuation may also impact the observed morphology, but given the spatial scale, this would then likely imply that different components are attenuated differently. The components are highly blended at the resolution of the MUSE data we explored here. The UV colours of the various components appear similar ($\Delta\beta<\pm0.3$), but they are challenging to measure as all combinations of ACS filters (F775W, F814W, and F850LP) are contaminated by Lyman-$\alpha$ emission and span a limited wavelength range. The data observed in IR filters (F105W, F125W, and F160W), on the other hand, were observed with WFC3, which has a poorer resolution. Furthermore, the IR filters are contaminated by a background LAE at $z=6.6$ slightly to the north-west of ID53. This significantly complicates the use of these data, for example, to obtain spatially resolved SED fits. 

As shown in Fig. $\ref{fig:HST}$, the MUSE data show spatial offsets between the UV continuum and the C{\sc iv}$_{1548}$ line emission, which we discuss in detail in \S $\ref{sec:CIVlines}$. No other spatial variations are detected in the MUSE data, that is, the spatial peak of the UV continuum does not shift significantly with wavelength, nor do any of the other detected emission features.

\subsection{Systemic redshift} \label{sec:zsys}
\begin{figure}
        \includegraphics[width=9.1cm]{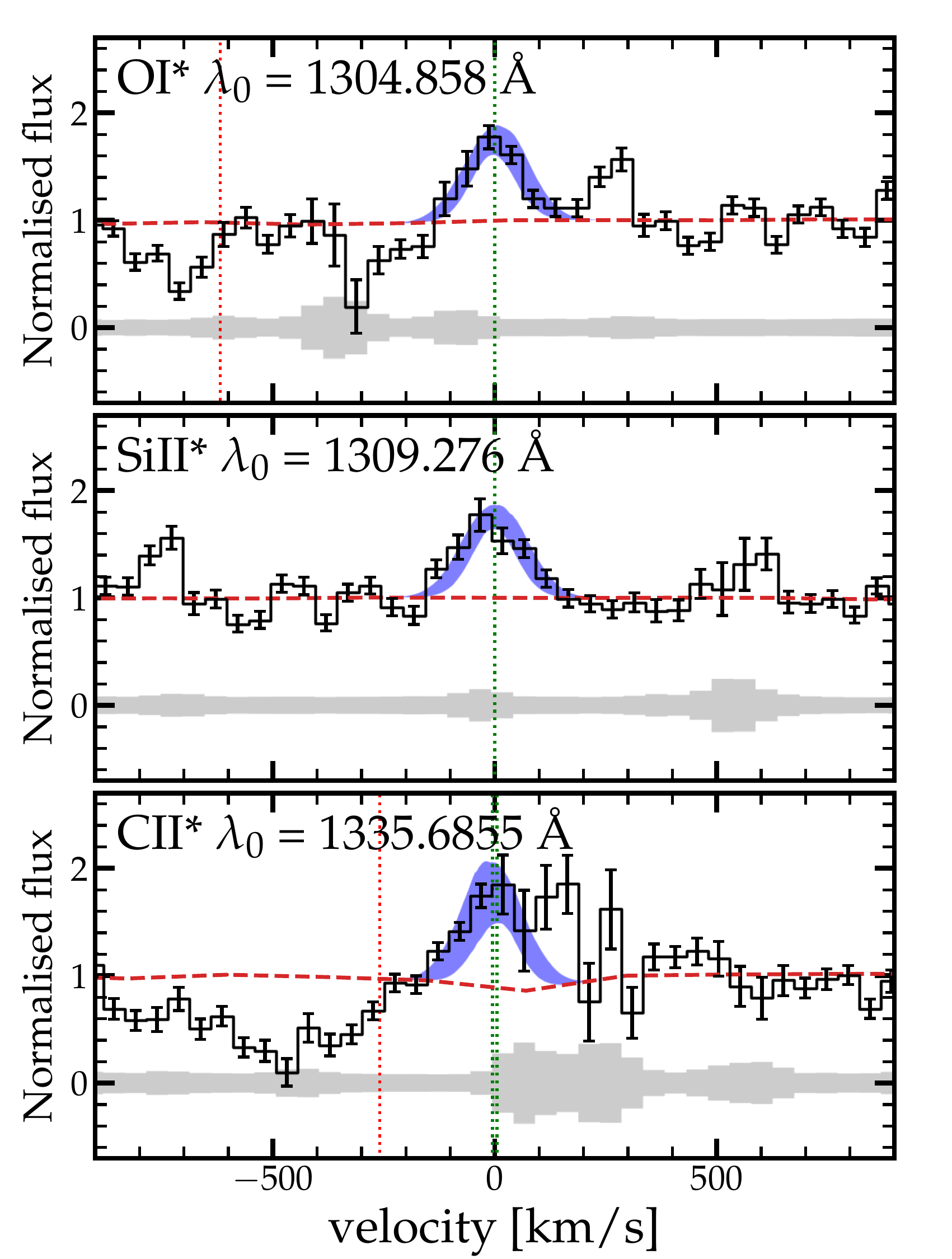} 
    \caption{Non-resonant fine-structure emission lines detected in ID53 (except for Si{\sc ii}$^*_{1533.43}$ , which is not shown). Vertical green lines mark the positions of the emission lines, and red lines mark the expected positions of absorption lines at their rest-frame velocity. The blue shaded region shows the fitted Gaussian profiles of the emission lines and their uncertainties. The dashed red line shows the underlying stellar continuum from the best-fit model with a single-burst SFH. We note that skyline contamination impacts the C{\sc ii}* line profile.}
    \label{fig:emission}
\end{figure}

The Ly$\alpha$ emission line of ID53 is clearly detected in the data. It appears as a single peak that is sharply skewed towards the red (Fig. $\ref{fig:1dspec}$), which is a well-known radiative transfer effect and typical for high-redshift Ly$\alpha$ emitters \citep[e.g.][]{Verhamme2006,Gronke2017}. The peak velocity of the red Ly$\alpha$ line is typically (red-) shifted with respect to the systemic redshift. The sensitivity of our spectrum allowed us to also detect the non-resonant fine-structure lines O{\sc i}*$_{1304}$, Si{\sc ii}*$_{1309}$, C{\sc ii}*$_{1336}$ , and Si{\sc ii}*$_{1533}$(Fig. $\ref{fig:emission}$). While the origin of these lines is somewhat unclear \citep[e.g.][]{Erb2010,Jaskot2014,Bosman2019}, they are empirically found to trace the systemic redshift in galaxies at $z\approx2-3$ \citep[e.g.][]{Shapley2003,Erb2010,Steidel2016,Vanzella2018,Marques2020}. Wind models from \cite{Prochaska2011} indicate that the detection of non-resonant emission suggests that there is little attenuation and that outflows only partially cover the source, particularly as we find that the spatial profiles of the non-resonant lines are similar to that of the UV continuum. More recently, \cite{Mauerhofer2021} showed a broad diversity of fluorescent line
profiles based on radiation hydrodynamical simulations of galaxy formation, but that they are usually centred on the systemic redshift. \cite{Smit2017b} empirically reported emission of the O{\sc i}*$_{1304}$ line with a redshift that is consistent with the redshift from the rest-frame optical [O{\sc ii}] doublet in an LAE at $z\approx5$. 

The spectral resolution of the MUSE data is sufficient to separate the emission of the non-resonant lines from resonant absorption lines, in particular because the absorption is found to be blue-shifted by typically $240$ km s$^{-1}$ (see Fig. $\ref{fig:emission}$). We fitted the non-resonant emission lines with a single Gaussian, where we allowed the central redshift and the normalisation to vary, but we required the width of the different lines to be the same. We subtracted the best-fit continuum model (a model with a single burst of star formation) before performing the fit using the {\sc lmfit} package for {\sc Python} \citep{Lmfit}. The choice of specific continuum model does not impact the measurement of the systemic redshift. The measured line width FWHM is $140\pm20$ km s$^{-1}$ and the best-fit systemic redshift is $z_{\rm sys} = 4.7745\pm0.0004$. The redshifts of the various lines agree perfectly. The systemic redshift implies that the red peak of the Ly$\alpha$ line is offset by $\Delta v_{\rm Ly\alpha} = +195\pm20$ km s$^{-1}$, see Table $\ref{tab:emlines}$. This is a typical value for galaxies with comparable Ly$\alpha$ EW \citep[e.g.][]{Adelberger2003,Nakajima2018b}.

\begin{table}
\caption{Detected emission lines.} 
\begin{tabular}{lrrr}\label{tab:emlines}
Emission-line & EW$_0$  & FWHM & $\Delta v$ \\  
 & [{\AA}] &  [km s$^{-1}$] &  [km s$^{-1}$] \\ \hline
Ly$\alpha$ & $62\pm3$ & $313\pm24$ & $+195\pm20$ \\
O{\sc i}$^*_{1304.86}$ & $0.56\pm0.09$  & $140\pm20$ & $1\pm18$\\
Si{\sc ii}$^*_{1309.28}$ & $0.56\pm0.08$ & $140\pm20$ & $-1\pm20$ \\
C{\sc ii}$^*_{1335.69}$ & $0.68\pm0.20$ & $140\pm20$ & $0\pm17$ \\
Si{\sc ii}$^*_{1533.43}$ & $0.66\pm0.19$ & $140\pm20$ & $-2\pm21$ \\
C{\sc iv}$_{1548.19}$ & $2.7^{+0.4}_{-0.4}$ & $110\pm40$ & $+51\pm19$  \\
C{\sc iv}$_{1550.77}$ & $1.3^{+0.6}_{-0.6}$ & $110\pm40$ & $+51\pm19$  \\  \hline
\multicolumn{4}{p{.45\textwidth}}{\footnotesize Properties of the emission lines detected in the integrated MUSE spectrum of ID53. $\Delta v$ lists the velocity offset with respect to the systemic redshift of $z_{\rm sys} = 4.7745\pm0.0004$. The Ly$\alpha$ flux and width are obtained from \cite{Leclercq2017}.}

\end{tabular} 
\end{table}

\begin{figure}
        \includegraphics[width=9.3cm]{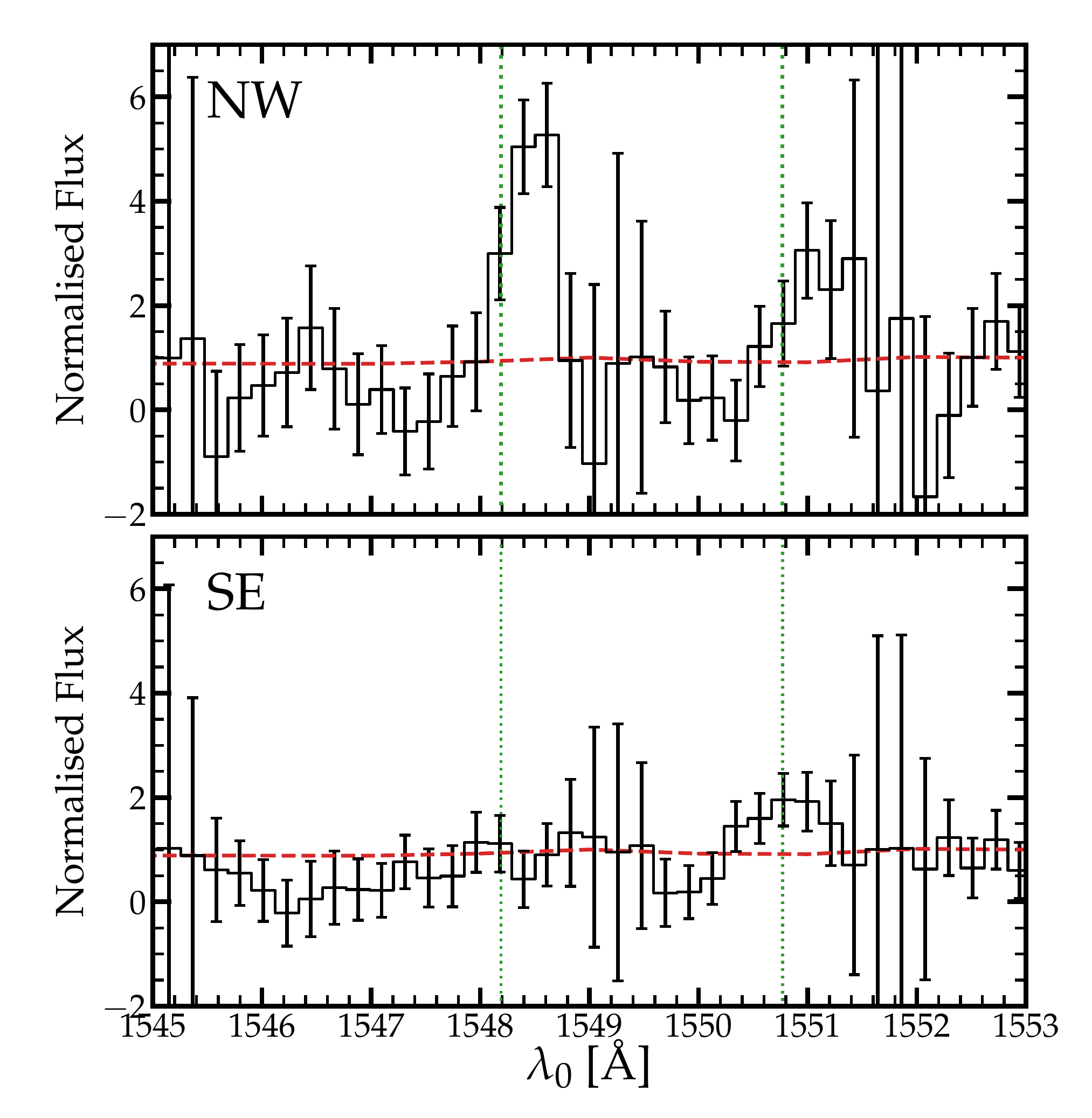}
    \caption{One-dimensional spectral extractions centred on the north-western (top) and south-eastern (bottom) part of ID53, focussing on the C{\sc iv}$_{1548,1550}$ doublet. The extraction is based on {\it HST}/ACS morphology (Fig. $\ref{fig:HST}$) using the {\sc Odhin} software. The dashed red line shows the continuum level from the best-fit stellar population model (see \S $\ref{sec:stellar_model}$). Spectra are normalised to this continuum level. Vertical dashed lines mark the expected position of the C{\sc iv} doublet at the systemic redshift. }
    \label{fig:carbon_resolved}
\end{figure}

\subsection{CIV emission} \label{sec:CIVlines}
The spectrum of ID53 clearly reveals narrow nebular C{\sc iv}$_{1548, 1550}$ emission lines; see Fig. $\ref{fig:1dspec}$. Because of the high ionisation energy of 47.9 eV, the detection of C{\sc iv} in emission suggests the presence of very hard ionising sources. C{\sc iv} is challenging to interpret because of the nearby interstellar absorption and complex stellar continuum \citep{Crowther2006,Leitherer2011,VidalGarcia2017}. Detailed investigation revealed that the C{\sc iv} line is stronger in the north-western part of the galaxy, with the spatial peak of the C{\sc iv}$_{1550}$ line coinciding with the fainter component of the galaxy seen in the {\it HST}/ACS imaging (Fig. $\ref{fig:HST}$). 

\begin{figure}
        \includegraphics[width=9.3cm]{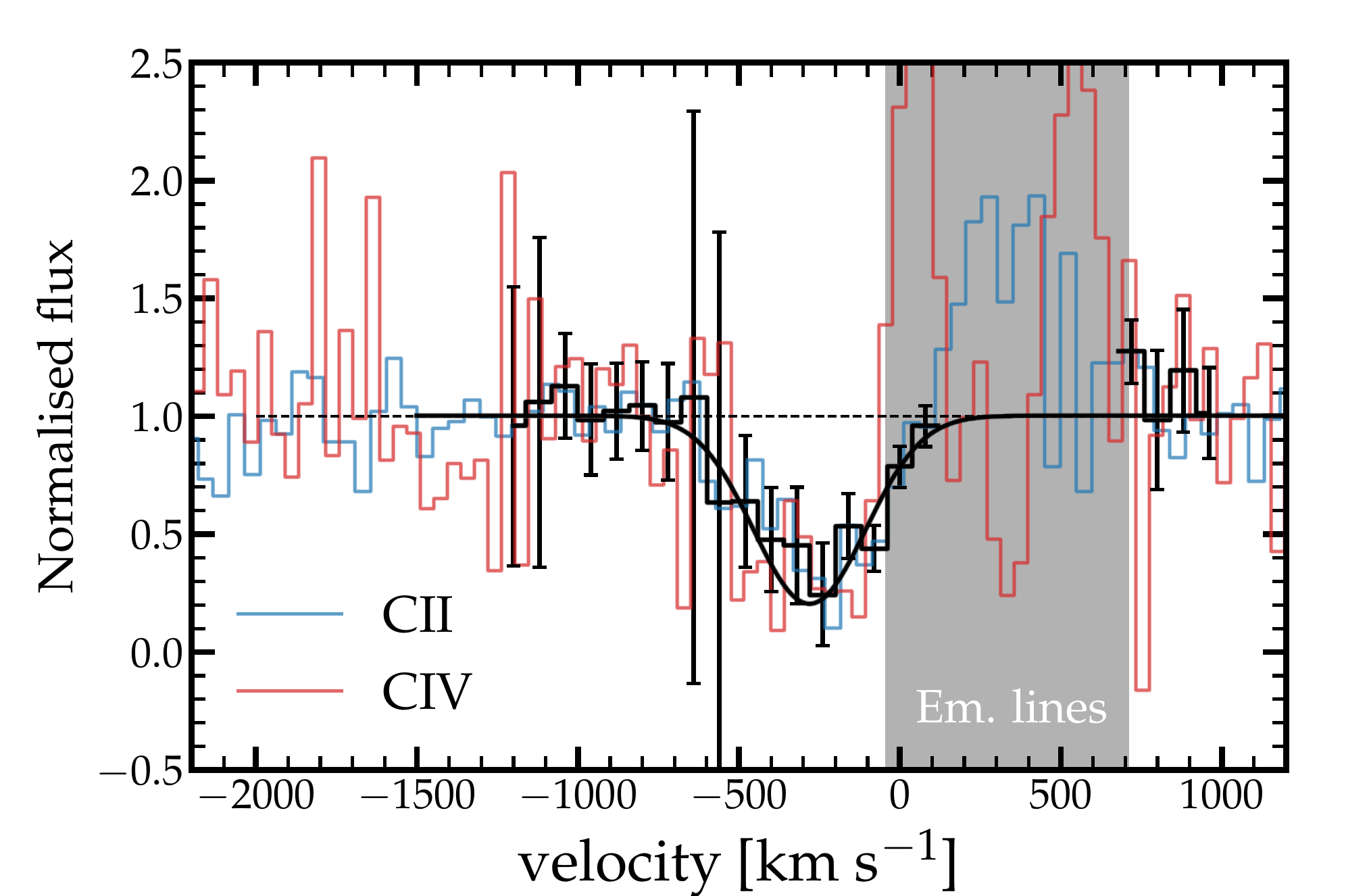} \\
        \includegraphics[width=9.3cm]{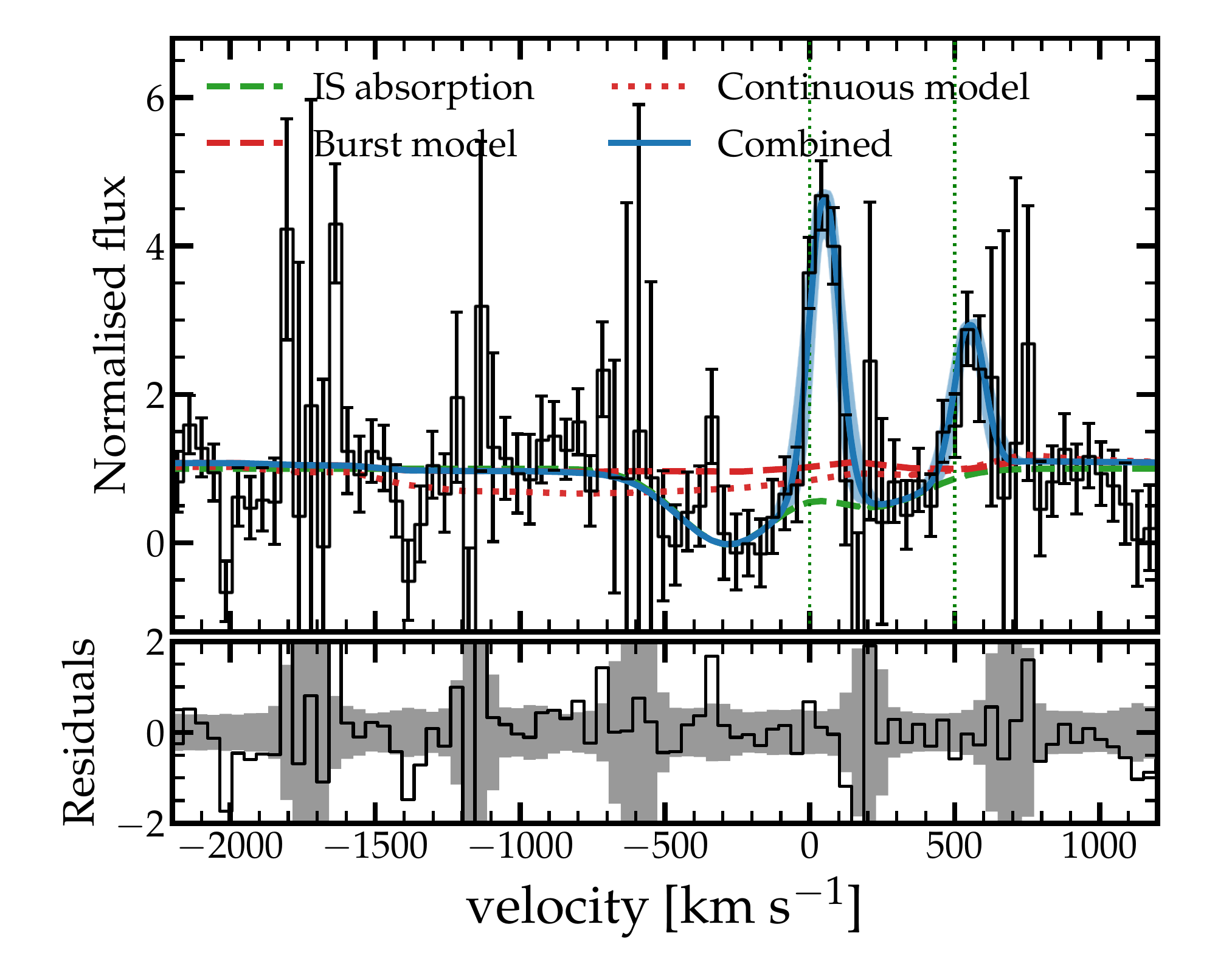} 
    \caption{Detailed spectrum of the C{\sc iv} line emission in ID53. Top: C{\sc ii}$_{1334}$ (blue) and C{\sc iv}$_{1548}$ (red) absorption in the integrated spectrum of ID53. The black data show the inverse-variance-weighted average of the two absorption transitions masking emission lines. The solid black line shows the best-fit Gaussian absorption profile. The grey shaded region highlights the position of the C{\sc iv} line that has been masked when measuring the weighted averaged spectrum. Bottom: C{\sc iv}$_{1548,1550}$ emission lines and the nearby continuum and absorption features. The velocity axes is plotted with respect to the rest-frame velocity of the C{\sc iv}$_{1548}$ line. The blue line shows the combined model of absorption, continuum, and emission, and its uncertainties. The combined model consists of stellar continuum (dashed red line), interstellar absorption (dashed green line), and nebular emission (blue shaded areas mark the 1$\sigma$ confidence interval). The small panel at the bottom shows the residuals between the combined model and data. }
    \label{fig:carbon_absorption}
\end{figure}

In Fig. $\ref{fig:carbon_resolved}$ we show the C{\sc iv} line in 1D spectral extractions centred on these distinct parts of the galaxy. These extractions are slightly blended and are therefore not fully independent of each other. In the NW, the C{\sc iv} lines are redshifted by $60\pm20$ km s$^{-1}$ with respect to the systemic redshift. In the SE, only a weak C{\sc iv}$_{1550}$ line is detected at the systemic redshift. In the integrated spectrum that is shown in the bottom panel of Fig. $\ref{fig:carbon_absorption}$, we also see evidence for blue-shifted C{\sc iv} absorption with a similar velocity and depth as the C{\sc ii} absorption. There is no strong stellar P-Cygni profile (discussed in more detail in \S $\ref{sec:specfit}$). We interpret this behaviour as follows: both components emit nebular C{\sc iv} emission, but the NW component has a significantly higher EW. Interstellar absorption is particularly present in the brighter component. The blue-shifted absorption in the 1550 transition thus renders the C{\sc iv}$_{1548}$ emission invisible in the SE. This explains why the spatial peak of the C{\sc iv}$_{1548}$ line coincides with the fainter component, while the position of the C{\sc iv}$_{1550}$ line does not (Fig. $\ref{fig:HST}$). 

The slight redshift of the C{\sc iv} lines in the NE with respect to the systemic redshift can be an intrinsic velocity shift due to an ongoing merger. It might also be due to radiative transfer effects, however \citep[e.g.][]{VillarMartin1996,Berg2019}. In analogy to well-studied Lyman-$\alpha$ radiative transfer \citep[e.g.][]{Verhamme2006}, scattering through an outflowing medium would yield double-peaked C{\sc iv} profiles with a dominant red peak. As the velocity offset between these peaks can be relatively small \citep[e.g.][]{Berg2019}, it is possible that the peaks are unresolved with the S/N and resolution of our data. However, such double peaks would be centred around the systemic velocity, suggesting that (even in the case of unresolved double peaks) there is an intrinsic velocity difference between the NE component and the systemic redshift (as measured by the non-resonant fine-structure lines).

We first attempted to measure the integrated C{\sc iv} EW of ID53 by controlling for interstellar absorption. Our method is motivated by the resemblance of the interstellar absorption profile of C{\sc iv}$_{1548}$ and the profile seen in C{\sc ii}$_{1334}$, which is less affected by nearby emission-infilling (top panel of Fig. $\ref{fig:carbon_absorption}$). We note that this resemblance could be a coincidence as the transitions do not trace the same gas phase. For C{\sc iv}$_{1548}$ we masked the $-50<v<750$ km s$^{-1}$ range, and for CII$_{1334}$ we masked $100<v<600$ km s$^{-1}$. Then we computed the inverse-variance-weighted average of the two absorption profiles and fitted this profile with a Gaussian using {\sc lmfit}. We find that the FWHM of the absorption profile is 164 km s$^{-1}$ and the centroid is at $-244$ km s$^{-1}$. The maximum absorption depth is 0.95 times the continuum. For the continuum, we used the best-fit stellar population model described in \S $\ref{sec:stellar_model}$. We then measured the C{\sc iv} luminosity by fitting Gaussian emission lines and correcting the continuum for this absorption profile for both lines, accounting for weaker absorption in the 1550 line due to the difference of a factor two in oscillator strength. The results are shown in the bottom panel of Fig. $\ref{fig:carbon_absorption}$. The fitted emission lines were forced to have the same width and central velocity. The resulting C{\sc iv} EWs are listed in Table $\ref{tab:emlines}$. 

In the integrated spectrum, the C{\sc iv} line is redshifted with respect to the systemic redshift by $51\pm19$ km s$^{-1}$ and it has a line width FWHM $119\pm40$ km s$^{-1}$. The correction for interstellar absorption mostly affects the C{\sc iv}$_{1548}$ line and enhances the EW by $+1.5$ {\AA}. The assumed SFH used for the stellar continuum model is of minor importance and only influences the EWs of both lines by $\pm0.1$ {\AA}. We combined the two IS absorption-corrected measurements to estimate an average combined C{\sc iv} EW of $4.0\pm0.7$ {\AA}. This EW is much higher than expected for pure stellar P-Cygni emission and is therefore predominantly of nebular origin.

Secondly, we attempted resolved measurements of the C{\sc iv} EW for the NW and SE parts of the galaxy. As their spectra have lower S/N and interstellar absorption may vary spatially (as argued above), for simplicity, we did not correct for absorption. In the NW we measure C{\sc iv}$_{1548,1550}$ EW of $2.1\pm0.4, 1.0\pm0.5$ {\AA}. This is consistent with the expected line luminosity ratio due to their relative oscillator strengths for an optically thin medium and thus indicates little scattering and interstellar absorption. In the SE we measure a C{\sc iv}$_{1548,1550}$ EW of $-0.1\pm0.2, 0.7\pm0.3$ {\AA,} which suggests significant impact of interstellar absorption. The C{\sc iv} EW of ID53 as a whole and of the NW part is comparable to the EW seen in several local low-metallicity dwarf galaxies with 12+log(O/H) $\lesssim 8.0$ \citep{Berg2016,Senchyna2017,Senchyna2019} and low-mass ($\approx10^{8-9}$ M$_{\odot}$) galaxies at $z\sim2-3$ \citep{Nakajima2018,Feltre2020}.

\begin{figure*}
        \includegraphics[width=18.7cm]{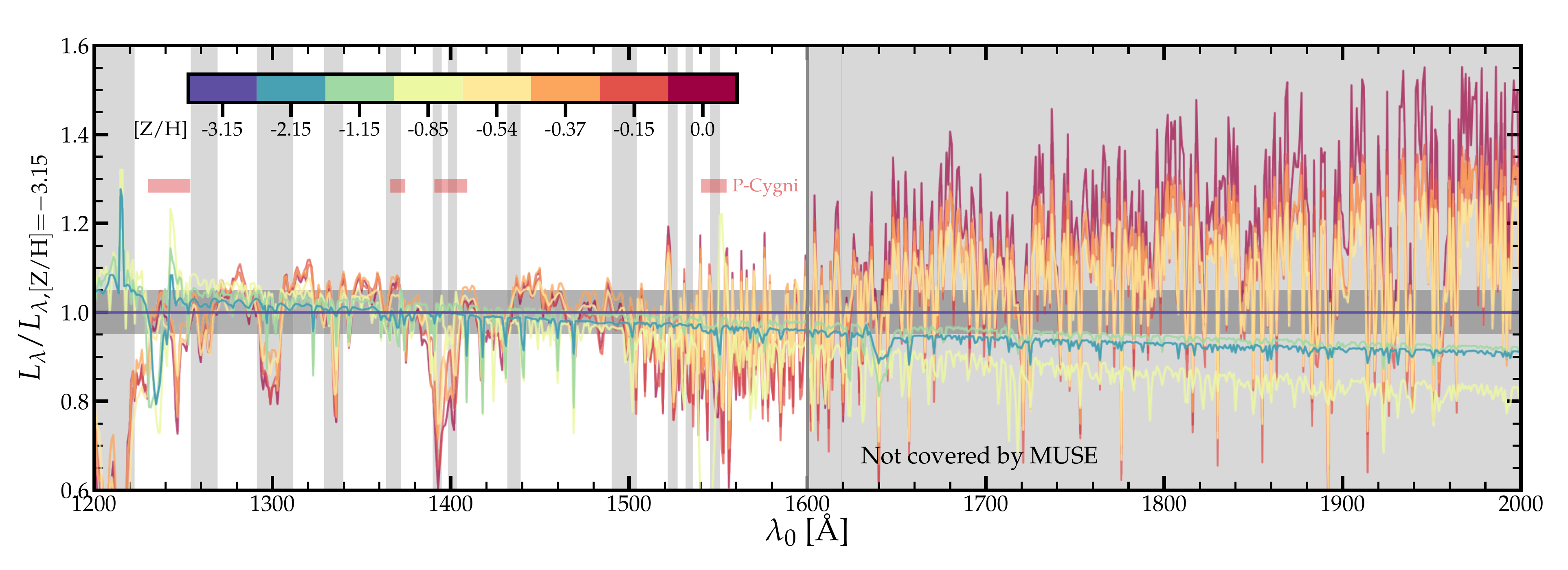} \\
    \caption{Constraining power of various features in terms of measuring the stellar metallicity of ID53. The vertical grey bands show wavelength regions that are not included in our analysis due to contamination from interstellar absorption or nebular emission (at $\lambda_0<1600$ {\AA}) or because they are not covered by the MUSE data ($\lambda_0>1600$ {\AA}). The coloured lines show the best-fit burst model for each stellar metallicity relative to the lowest-metallicity model. The models are fit to the MUSE data only, ignoring wavelengths with features from interstellar lines or nebular emission. The horizontal grey band illustrates the typical S/N of our data. We illustrate the locations of the N{\sc v}, O{\sc v}, Si{\sc iv,} and C{\sc iv} P-Cygni features in red.}
    \label{fig:photospheric}
\end{figure*}

\section{Stellar population modelling} \label{sec:stellar_model}
The UV continuum over the $\lambda=1200-2600$ {\AA} wavelength range contains a plethora of stellar features. The most prominent features in young stellar populations are stellar wind features such as N{\sc v} and C{\sc iv} \citep[e.g.][]{Leitherer2001,Steidel2016} and photospheric line-blanketing that is mostly seen in the $\lambda=1600-2600$ {\AA} range \citep[e.g.][]{Rix2004}; see Fig. $\ref{fig:photospheric}$. The strengths of these features in integrated galaxy spectra are sensitive to the stellar metallicity, but also to the present-day mass distribution functions of stars (i.e. the SFH and the IMF). Stellar features can be blended with interstellar absorption lines and nebular emission. Additionally, the observed spectrum in the UV continuum may be affected by dust attenuation and nebular continuum emission. However, these last two effects mostly affect the normalisation and the slope of the continuum, unlike the metallicity, which predominantly affects the ability of the spectrum to wiggle (see also \citealt{Cullen2019}).

In this section we analyse the observed UV spectrum of ID53. The MUSE data spans $\lambda_0=900-1600$ {\AA}. We first explore the full spectral information and then focus on the N{\sc v} and C{\sc iv} P-Cygni wind features. These winds originate when gas is pushed away from massive stars by radiation pressure \citep{Castor1975}. Their profile depends on the ionisation structure, velocity, and mass-outflow rate \citep[e.g.][]{Lamers1993}. We explore models where the star formation occurred in a single burst and models with a constant star formation history. We only use wavelength regions in the MUSE spectra that are unaffected by absorption or emission in the interstellar or the inter-galactic medium (i.e. we use wavelengths between $1216$ and $1600$ {\AA} from mask 1 defined in Table 3 of \citealt{Steidel2016}). We fit primarily to the MUSE data, but we use photometry from {\it HST}/WFC3 and {\it Spitzer}/IRAC as bounds.

\subsection{Modelling considerations} \label{sec:considerations}
Our composite stellar population models are based on single stellar populations from BPASS (v2.2; \citealt{BPASS2018}), which are the models that match the spectrum of typical Lyman-break galaxies at $z\approx2$ best \citep{Steidel2016}. We combined these models with a \cite{Reddy2016} dust attenuation law. We simulated a large grid of models with ages from log$_{10}$(age/yr) = 6.0...8.2 in steps of 0.1 dex and stellar metallicities $Z=10^{-5}, 10^{-4},$ 0.001, 0.002, 0.004, 0.006, 0.008, 0.01, 0.014, 0.02, and 0.03 (i.e. [Z/H]=-3.15, -2.15, -1.15, -0.85, -0.54, -0.37, -0.24, -0.15, 0.00, 0.15, and 0.33 for the solar metallicity assumed here). These ages correspond either to the age of the burst or to the time since the onset of the continuous star formation. In the latter case, the metallicity is the same throughout the history. We varied E($B-V$) from 0.0 to 0.5 in steps of 0.02, and we simulated models with initial stellar masses $10^7 - 10^{11}$ M$_{\odot}$ in steps of 0.05 dex. In practice, we find little degeneracy between the dust attenuation, metallicity, and stellar age because we find that the slope of the UV continuum is only weakly dependent on metallicity and stellar age in the range included here. We used a standard \cite{Chabrier2003} IMF with mass limits 1.0-100 M$_{\odot}$ in this analysis. In Appendix $\ref{sec:IMFvar}$ we show that the results are not strongly sensitive to the specific choice of the IMF. The inclusion of binary stars improves the fit, but does not significantly impact the best-fit parameters.

We did not include nebular emission or interstellar absorption in our models. This means that we may slightly overestimate the dust attenuation (as nebular continuum emission may contribute to the reddening of the spectrum regardless of dust). This limitation of our models is mostly important when comparing to the longer wavelength photometry. The H$\alpha$ line influences the [3.6] flux (e.g. \citealt{Raiter2010IRAC,Bouwens2016Xion,Maseda2020}, and Section $\ref{sec:Halpha}$). The $K_s$ band contains the [O{\sc ii}] doublet and the Balmer break, but this wavelength region is also mostly affected by nebular continuum emission \citep{Raiter2010}. Nebular continuum emission can contribute up to $\approx40$ \% of the total flux around the Balmer jump in case of SSPs with ages $<3\times10^6$ yr \citep{Reines2010,Byler2017,Gunawardhana2020}. Similarly, the {\it Spitzer}/IRAC [4.5] band can be significantly influenced by nebular continuum emission that is strong around the Paschen jump at $\lambda_0=8207$ {\AA}. Therefore, we used the photometric information in the IR as lower bounds (i.e. the attenuated stellar light cannot be brighter than the observed magnitudes).

\begin{figure*}
\centering
        \includegraphics[width=18.3cm]{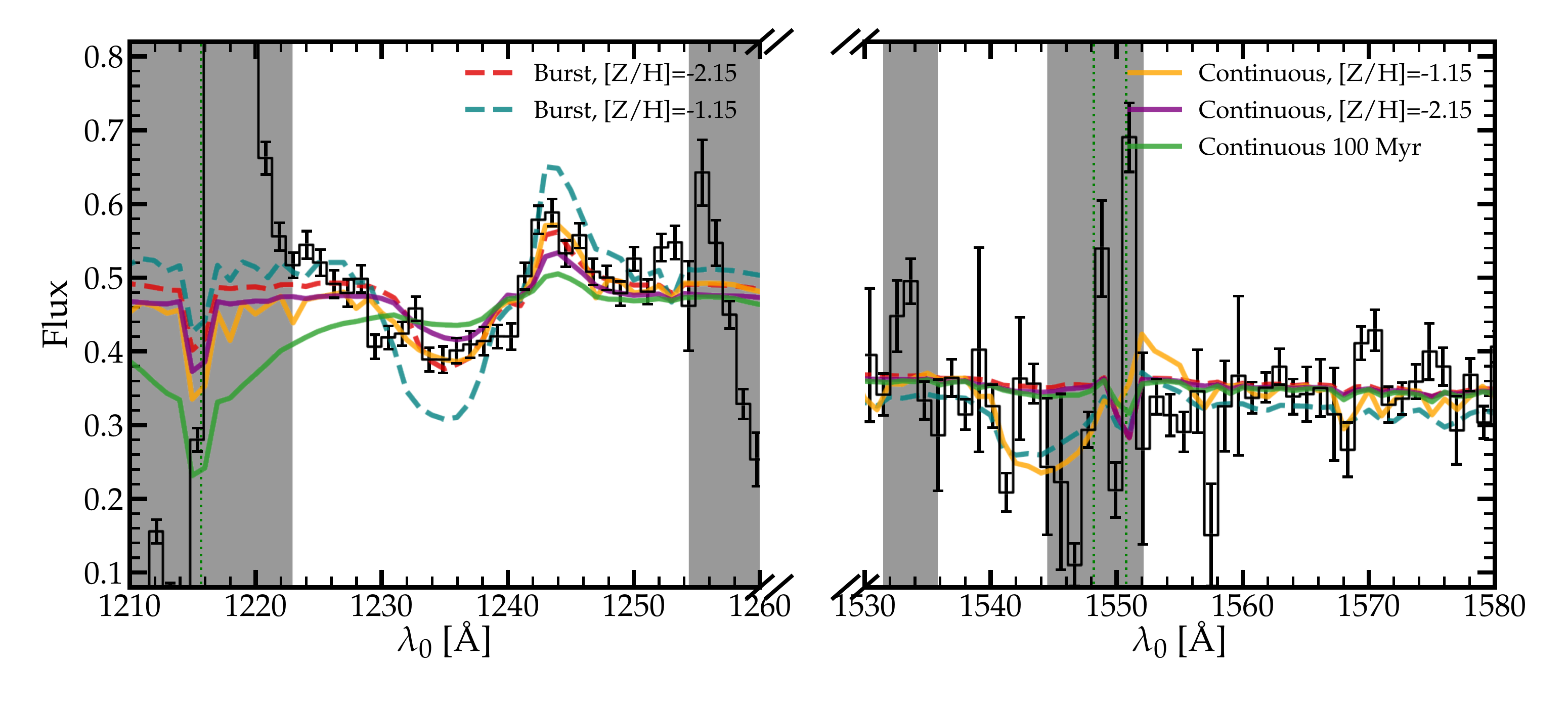} 
    \caption{Spectrum of ID53, zoomed-in on the rest-frame wavelengths surrounding the stellar P-Cygni features from N{\sc v} (left) and C{\sc iv} (right). A N{\sc v} P-Cygni feature is clearly detected. No significant P-Cygni profile from C{\sc iv} emission is detected in either absorption at $\lambda_0\approx1542$ {\AA} or in emission at $\lambda_0\approx1551$ {\AA}. The grey shaded regions of the spectrum are not included in the fitting as they are contaminated by nebular emission or interstellar absorption. We show the best-fit models with a single-burst and continuous star formation for a metallicity [Z/H]=-2.15 and [Z/H]=-1.15. These are the best-fit metallicities for the single-burst and continuous star formation, respectively, and are first listed in the legends. For reference, we also show the best-fit model with continuous star formation that is ongoing for 100 Myr in green, which has a best-fit [Z/H]=-2.15.} 
    \label{fig:Pcyg}
\end{figure*}

Previous studies that constrained stellar metallicity in high-redshift galaxies based on spectral fitting in the rest-frame UV have typically used constant SFHs with a fixed age of $10^8$ yr \citep[e.g.][]{Steidel2016,Cullen2019}. This is the equilibrium timescale after which the UV continuum does not change much for a significant amount of time. This assumption is likely valid for these studies, as stacking tends to smooth bursty star formation histories as long as the bursts are not in phase \citep[e.g.][]{MattheeSchaye2019}.  \cite{Cullen2019} verified this assumption using average SFHs based on hydrodynamical simulations. As the galaxies in these studies were selected by their UV continuum luminosity, it is also plausible that there is no particular preference for selecting galaxies experiencing a short-lived burst. Furthermore, in an analysis of individual galaxies with M$_{\rm star}\approx10^{10}$ M$_{\odot}$ at $z\approx2$, \cite{Topping2020b} used a constant SFH with a lower age limit of $10^7$ yr motivated by the dynamical timescale and typical H$\alpha$ EW of the sample, which validates these ages.  

Unlike these studies, our target is a galaxy with a lower stellar mass of $\approx10^9$ M$_{\odot}$ , and it may be caught in a moment in which it experiences a starburst, as indicated by the high Ly$\alpha$ EW. Therefore, we cannot use the assumption of a constant SFH on the equilibrium timescale of the UV continuum or the dynamical timescale, and we have to vary the star formation history. This means that we included populations of stars with ages younger than $10^7$ yr, which significantly complicated the fitting process (see e.g. \citealt{Chisholm2019}).

\subsection{Full spectral fitting} \label{sec:specfit}   
In order to illustrate the constraining power of the features in the observed part of the UV spectrum, we show in Fig. $\ref{fig:photospheric}$ the best-fit single burst models for each metallicity from [Z/H]=-2.15 to [Z/H]=0.0 relative to the best-fit [Z/H]=-3.15 model. We only used the MUSE data (masking interstellar and nebular features) in the fitting procedure. Fig. $\ref{fig:photospheric}$ illustrates that metal-sensitive features are typically more prominent and offer stronger constraining power at $\lambda_0=1600-2000$ {\AA}, in particular in the regime [Z/H]$\gtrsim-1$ \citep[see also][for a discussion of specific wavelength intervals that are particularly metal sensitive]{Sommariva2012,VidalGarcia2017}. These features are unfortunately not covered by the MUSE data. Additionally, stellar wind features around N{\sc v} ($\lambda\approx1240$ {\AA}), Si{\sc iv} ($\lambda\approx1400$ {\AA}), and C{\sc iv} ($\lambda\approx1550$ {\AA}) are strongly metal sensitive in the lower-metallicity regime, although in the latter two cases this is also complicated by interstellar absorption and emission. 

Fig. $\ref{fig:photospheric}$ also illustrates that higher-metallicity models tend to be brighter in redder wavelengths. This suggests that including the photometric data in the NIR from {\it HST}/WFC3 and {\it Spitzer}/IRAC can constrain higher metallicity models. We find that the criterion that the stellar light in the F105W and [4.5] filters cannot be brighter than the observed magnitudes offers useful constraining power. We therefore combined the full spectroscopic data of the stellar continuum of ID53 with long-wavelength photometry. Specifically, we calculated $\chi^2_{\rm red}$ over wavelength regions in the MUSE data and added the bounds [4.5]$>25.3$ and F105W$>25.2$. Our fitting results for both single-burst (i.e. a single stellar population) and continuous star formation are summarised in Table $\ref{tab:chisel}$. The best-fit models for each metallicity are listed in Table $\ref{tab:chisel_complete}$.

\begin{table*}
\centering
\caption{Results of the SED modelling.} \label{tab:chisel}
\begin{tabular}{lrrrrrr} \hline

Type SFH  & [Z/H]$_{\rm full}$ & log$_{10}$(Age/yr) & log$_{10}$(M$_{\rm star}$/M$_{\odot}$) & log$_{10}$(L$_{\rm H\alpha, int}$/erg s$^{-1}$) & $\chi^2_{\rm red}$  \\ 

Burst, bin, Chab100  & $-2.15^{+1.15}_{-1.0}$  & $6.5^{+0.6}_{-0.3}$ & $8.6^{+0.7}_{-0.2}$ & $43.6^{+0.3}_{-1.1}$ & 3.00   \\
Continuous, bin, Chab100  &  $-1.15^{+0.45}_{-2.0}$  & $7.3^{+0.6}_{-0.8}$ & $9.0^{+0.3}_{-0.4}$ & $43.5^{+0.4}_{-0.4}$ & 3.17   \\
\hline
\multicolumn{7}{p{.81\textwidth}}{\footnotesize The uncertainties correspond to the 1$\sigma$ intervals that are estimated following the standard method described in e.g. \citealt{Avni1976} assuming that the likelihood scales with $\propto \exp(-\chi_{\rm red}^2/2)$ The $\chi^2_{\rm red}$ computed over the full MUSE spectrum of the best-fit models are listed in the right-most column. The best-fit stellar attenuation is E$(B-V)=0.18^{+0.11}_{-0.06}$ for the burst and continuous models.}
\end{tabular}
\end{table*}

The grey horizontal band in Fig. $\ref{fig:photospheric}$ and the $\chi^2$ values in Table $\ref{tab:chisel_complete}$ show that the S/N of the MUSE data on their own is in general only sufficient to significantly distinguish models with [Z/H]$>-0.8$ from models with lower metallicities. It is challenging to distinguish models with [Z/H] between $-3.15$ to $-0.8$, which all provide reasonable fits to the data with $\Delta \chi^2_{\rm red} < 1$ with respect to the best fit. With this in mind, we focus our discussion of the results in the metal-poor regime on stellar wind features that offer most constraining power. Furthermore, we note that in \S $\ref{sec:nebular}$ we include constraints from Balmer lines that help distinguishing the model fits in this regime.

The N{\sc v} P-Cygni feature is clearly detected in a wavelength region that is free of skylines; see Fig. $\ref{fig:Pcyg}$. We find that a continuous SFH reproduces the N{\sc v} profile slightly better than single-burst models (full details are listed in Table $\ref{tab:chisel_complete}$). This is because the absorption part of the modelled feature is too narrow in the burst model, and it overpredicts the strength of the emission part.\footnote{We note that emission infilling from broad Ly$\alpha$ is unlikely as it would require significant flux at $\gtrsim+5000$ km s$^{-1}$.} This can be seen by comparing the dashed red model (the best-fit burst) to the yellow model (the best-fit continuous model) in Fig. $\ref{fig:Pcyg}$ and the $\chi^2$ difference listed in Table $\ref{tab:chisel_complete},$ which corresponds to a $\approx1\sigma$ difference. The N{\sc v} profile is best fit with models with stellar metallicity [Z/H]=-1.15 to -0.84. Higher-metallicity models yield an N{\sc v} profile that is too strong, while the N{\sc v} feature is too weak in the lowest-metallicity model that we considered. The continuous SFH models that best-match the N{\sc v} feature have ages of $2\times10^7$ yr. The best-fit burst model is much younger (because the N{\sc v} P-Cygni feature originates mostly from short-lived very massive stars), with an age of $3\times10^6$ yr \citep[e.g.][]{Chisholm2019}. In Fig. $\ref{fig:Pcyg}$ we also show the best-fit model with continuous star formation that is ongoing for 100 Myr as this is the standard SFH that is typically used, as discussed above. It is clear that this model (which has a best-fit metallicity [Z/H]=-2.15) does not match the spectrum around Ly$\alpha$ and the N{\sc v} P-Cygni profile well, and it is ruled out at $>1\sigma$ when comparing its $\chi^2_{\rm red}$ over the full spectral range and $>2\sigma$ when focussing on N{\sc v} alone.

The interpretation of the observed C{\sc iv} spectrum is complicated by interstellar C{\sc iv} absorption and nebular emission \citep[e.g.][]{VidalGarcia2017}, particularly in the range $\lambda_0 = 1547-1551$ {\AA}. Interstellar C{\sc iv}$_{1548}$ absorption is detected out to $\approx - 500$ km s$^{-1}$ and nebular emission (discussed in \S $\ref{sec:CIVlines}$) is also clearly seen, see Fig. $\ref{fig:carbon_absorption}$. These features are masked when fitting the stellar continuum models, as illustrated by the grey shaded regions in Fig. $\ref{fig:Pcyg}$. We do not see clear evidence of a stellar P-Cygni profile in the C{\sc iv} line, although the BPASS models predict it to be broader than the interstellar absorption. There appears to be faint absorption at $\approx-1500$ km s$^{-1}$ ($\lambda_0=1541$ {\AA}, most clearly seen in Fig. $\ref{fig:carbon_absorption}$) that may be of stellar origin, but we do not see possible stellar absorption around $-1000$ km s$^{-1}$ ($\lambda_0=1544$ {\AA}). There is no evidence for a broad emission component redwards of the 1550 {\AA} line. Because of these characteristics, the data around the C{\sc iv} feature prefer a metallicity of [Z/H]=-2.15 to -1.15 as higher metallicities would yield much stronger P-Cygni profiles.

In addition to N{\sc v} and C{\sc iv}, we have explored the presence of P-Cygni features from O{\sc v}$_{1371}$ and Si{\sc iv}$_{1400}$ in the data and the models. We find that the O{\sc v} line is in a wavelength region that is heavily affected by skyline residuals. SiIV is affected by interstellar absorption similarly to C{\sc iv}. Moreover, the variations in Si{\sc iv} and O{\sc v} between various models are smaller than variations in N{\sc v} and C{\sc iv}, meaning that they are less constraining.

As a consistency check, we also measured the metallicity by fitting specific metallicity-sensitive wavelength ranges. In our data, the 1425 {\AA} index \citep[e.g.][]{Sommariva2012,VidalGarcia2017} is in a wavelength region with good S/N and is not affected by interstellar emission or absorption. Although literature results show that this index is particularly sensitive to metallicity variations in the range in which metallicities are higher than $0.5$ Z$_{\odot}$ and ages are older than 30 Myr, we still find that it yields a metallicity of [Z/H]=$-1.15^{+0.5}_{-2.0}$ (single burst) and [Z/H]=$-1.15^{+0.45}_{-2.0}$ (continuous), which is well consistent with full spectral fitting.

\subsection{Summary of the model results}
In this section we have shown that most importantly, , detailed features in the rest-frame UV continuum spectrum of ID53 can be well matched by BPASS v2.2 models combined with a \cite{Chabrier2003} IMF with standard upper mass limits of 100 M$_{\odot}$ and a \cite{Reddy2016} attenuation law when ages younger than 100
Myr are allowed. Regardless of the assumed SFH, the rest-frame UV spectrum of ID53 combined with bounds on the IR photometry implies a best-fit stellar metallicity in the range [Z/H]=-2.1 to -1.1 (Table $\ref{tab:chisel}$). Metallicities higher than [Z/H]$>-0.84$ are firmly ruled out. As single-burst or continuous star formation models yield a similar match to the MUSE data, we combined the metallicity and mass likelihood distributions of these models to derive fiducial measurements of log$_{10}$(M$_{\rm star}$/M$_{\odot}$) = 8.6$^{+0.6}_{-0.2}$ and [Z/H]=$-2.15^{+1.25}_{-0.50}$.

%%%%
\section{Constraints from Balmer lines}  \label{sec:nebular}  
In order to improve our knowledge of the star formation histories that are crucial to know for the stellar population modelling of ID53, we required additional information beyond the stellar continuum. Recombination lines originating from H{\sc ii} regions, such as those from the Balmer and Lyman-series, are directly sensitive to the luminosity in the ionising ($\lambda<912$ {\AA}) part of the spectrum. This is particularly useful as we cannot directly observe this part of the spectrum. In addition to the ionising luminosity, the produced emission-line luminosities are sensitive to the electron temperature, the ionising escape fraction, and the attenuation within H{\sc ii} regions. We used tabulated H$\alpha$ luminosities from BPASS \citep{BPASS2018} that were calculated with {\sc Cloudy} \citep{Ferland2013}. We note that these luminosities correspond to a scenario in which the escape fraction and the dust absorption of ionising photons within the H{\sc ii} regions are negligible.

With the currently available data, we can particularly investigate the strength of the H$\alpha$ and Ly$\alpha$ emission lines that various stellar models produce. However, including these data introduces several caveats, in particular the dust attenuation of emission lines compared to the stellar continuum \citep[e.g.][]{Shivaei2020}, and the Ly$\alpha$ escape fraction \citep{Hayes2015}. Therefore, instead of using H$\alpha$ and Ly$\alpha$ measurements as bounds to the models, we compared their measured luminosities to the intrinsic line luminosities predicted by the models in order to infer the required attenuation and escape fraction, which enables a qualitative assessment of the validity of the various models.

\subsection{H$\alpha$} \label{sec:Halpha}
Observationally, we can infer the H$\alpha$ luminosity for ID53 because it has a dominant contribution to the flux in the [3.6] filter on {\it Spitzer}/IRAC at $z=4.77$ \citep[e.g.][]{Raiter2010IRAC,Shim2011,Stark2013,Harikane2018,Maseda2020}. At the redshift of ID53, the [4.5] filter is not contaminated by strong line emission and can serve as a continuum estimate. We observe a strong colour excess of [3.6]-[4.5]=$-0.6\pm0.1$ that is indicative of H$\alpha$ emission. The continuum slope around H$\alpha$ depends on the (attenuated) stellar population model and on the contribution of nebular continuum. We find that our attenuated stellar population models all have a continuum slope in the range $\beta=-2.2$ to $\beta=-2.6$ over $\lambda_0=6000 - 8000$ {\AA}. We did not use the stellar population models to estimate the continuum slope as nebular continuum emission can significantly flatten the slope of the integrated light. We therefore conservatively assumed that $\beta$ is in the range $\beta=0.0$ to $-2.6$, but list our inferred H$\alpha$ luminosity assuming a mean $\beta=-2.0$. We also assumed that excess line emission in the [3.6] filter comes purely from H$\alpha$ as [N{\sc ii}] is typically very weak ($<1$ \% of H$\alpha$) in LAEs at $z\approx2$ \citep[e.g.][]{Trainor2016,Matthee2021} and at low metallicities in general. We measure an H$\alpha$ luminosity L$_{\rm H\alpha}=3.9^{+1.5}_{-0.6}\times10^{42}$ erg s$^{-1}$ and a corresponding EW$_0 = 710^{+390}_{-250}$ {\AA} based on the photometry listed in Table $\ref{tab:photometry}$. The uncertainties were calculated by propagating the measurement uncertainties and allowing a range of continuum slopes $\beta=-2.6 - 0$.

\begin{figure}
\centering
        \includegraphics[width=9.3cm]{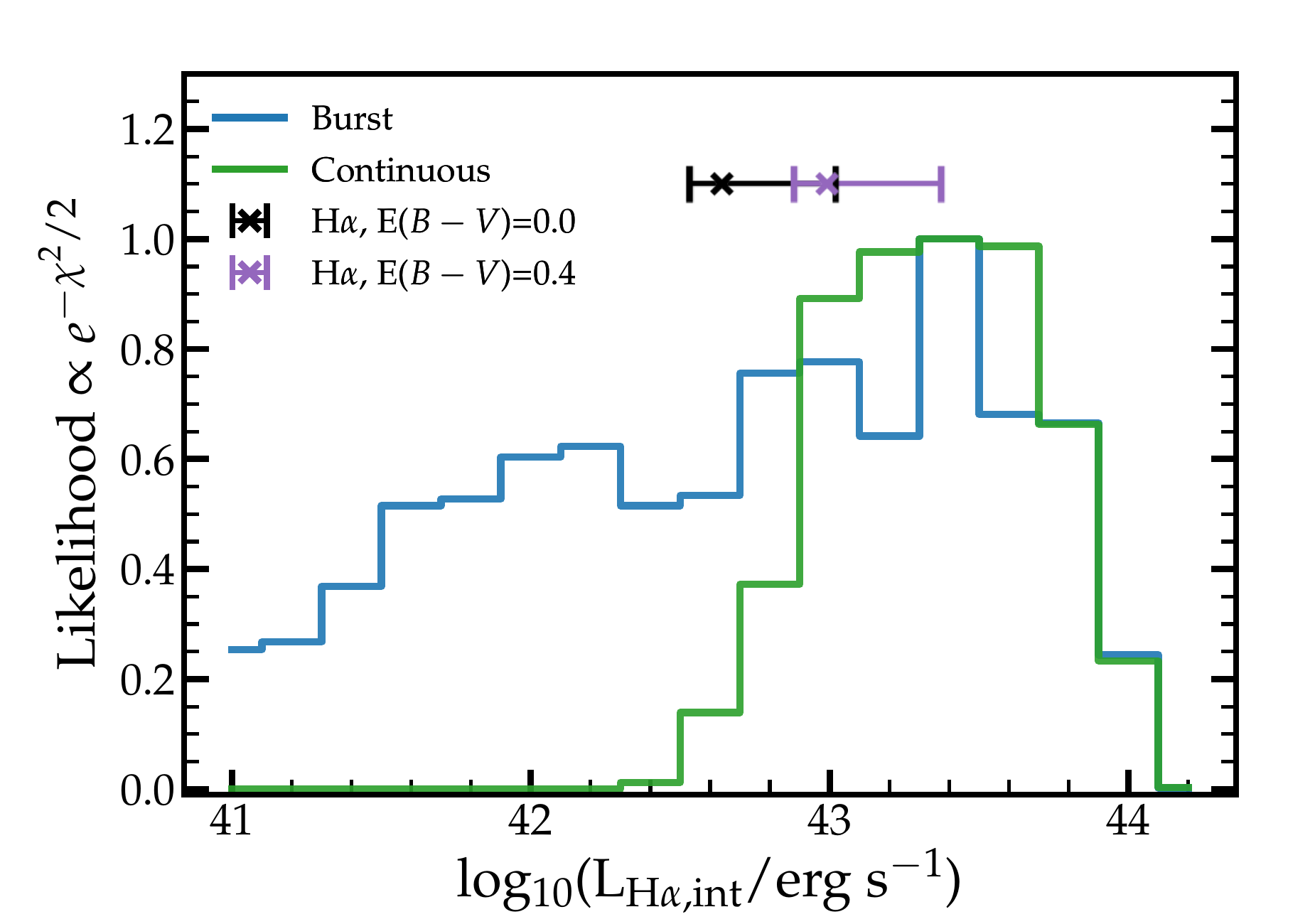} 
    \caption{Likelihood of the intrinsic H$\alpha$ luminosities associated with the stellar population models presented in \S $\ref{sec:stellar_model}$. The data points show the inferred H$\alpha$ luminosity from {\it Spitzer}/IRAC data when it is either unattenuated (black) or attenuated relatively significantly with E$(B-V)=0.4$ (i.e. A$_{\rm H\alpha}\approx1$).  }
    \label{fig:Halpha}
\end{figure}

The intrinsic nebular H$\alpha$ luminosities associated with the BPASS SSP models are tabulated \citep{BPASS2018} and are therefore known for each of our stellar models (Table $\ref{tab:chisel}$). However, comparison to the observations requires an estimate of the dust attenuation. We assumed that the nebular attenuation follows a \cite{Cardelli1989} law with $k_{\rm H\alpha} = 2.52$ \citep{Reddy2020}. Our SED fits suggest a stellar attenuation E$(B-V)\approx0.2$ (with a firm bound E$(B-V)<0.4$), but it is unclear whether the nebular lines are similarly attenuated in high-redshift galaxies \citep{Kashino2013,Theios2019,Shivaei2020}. 

When E($B-V$)$_{\rm neb}=0.4$, the observed H$\alpha$ luminosity implies an intrinsic luminosity of log$_{10}$(L$_{\rm H\alpha}$/erg s$^{-1}$)=$42.99^{+1.4}_{-0.5}$. For a similar nebular and stellar attenuation (E$(B-V)_{\rm neb}=0.2$), the expected intrinsic H$\alpha$ luminosity is log$_{10}$(L$_{\rm H\alpha}$/erg s$^{-1}$)=$42.79^{+1.4}_{-0.5}$. Remarkably, these H$\alpha$ luminosities are all well below the intrinsic H$\alpha$ luminosities for our best-fit stellar models that have a single-burst or continuous star formation; see Table $\ref{tab:chisel}$ and Fig. $\ref{fig:Halpha}$. For the best burst and continuous models, the intrinsic H$\alpha$ luminosity would only match the inferred luminosity in case the attenuation is as high as E$(B-V)_{\rm neb}=1.0,$ and a significant range of models with continuous star formation that fit the UV continuum reasonably well require even higher attenuation. 
 
Possible explanations of the large difference between the modelled and observed H$\alpha$ luminosity could be a more complex SFH (see \S $\ref{sec:discuss_complex}$), absorption of ionising photons within H{\sc ii} regions, or a high escape fraction of ionising photons that would reduce the H$\alpha$ luminosity \citep[e.g.][]{Zackrisson2017,Naidu2021}. Future detailed spectroscopic measurements of the attenuation and the H$\alpha$ luminosity, for example through NIRspec spectroscopy with the {\it James Webb Space Telescope (JWST)} will be very useful to help constrain the SFH and the stellar metallicity when they are combined with rest-frame UV spectroscopy. Such detailed studies will be enabled soon when various deep planned spectroscopic surveys in the field of ID53 are undertaken with {\it JWST}, such as JADES (PI Rieke/Ferruit) and FRESCO (PI Oesch). However, we stress that {\it JWST} will spatially resolve the resolved structure seen in the {\it HST} data, which may make observations through narrow slits challenging to interpret. Although some intrinsic uncertainty related to the conversion of the intrinsic H$\alpha$ luminosity into the nebular line luminosity will remain, we illustrate in Appendix $\ref{sec:ApC}$  that even 25 \% errors on the intrinsic H$\alpha$ luminosity already allow much more accurate stellar population modelling.

\subsection{Ly$\alpha$ escape fraction}
The Ly$\alpha$ escape fraction, $f_{\rm esc, Ly\alpha}$, can be calculated from the ratio of the observed Ly$\alpha$ luminosity to the intrinsic H$\alpha$ luminosity: $f_{\rm esc, Ly\alpha}= L_{\rm Ly\alpha}/(8.7 L_{\rm H\alpha, int})$ \citep[e.g.][]{Henry2015}. There are various estimates of the intrinsic H$\alpha$ luminosity: either through our SED models or through a (dust-corrected) H$\alpha$ luminosity that is inferred from the IRAC [3.6] excess. Our stellar population models imply $f_{\rm esc, Ly\alpha}= 14^{+90}_{-10}, 6^{+12}_{-4}$\% for single-burst and continuous SFHs, respectively.  From the inferred H$\alpha$ luminosity we measure a Ly$\alpha$ escape fraction $f_{\rm esc, Ly\alpha}=24^{+7}_{-5}$\% in case E($B-V$)=0.2. This would be in agreement with the expected escape fraction of 30\% based on an empirical relation between the Ly$\alpha$ EW and $f_{\rm esc, Ly\alpha}$ observed over $z=0-2$ \citep{SM2019}. Moreover, based on a similar {\it Spitzer}/IRAC-based methodology, \cite{Harikane2018} reported a median $f_{\rm esc, Ly\alpha}=27$\% for average LAEs at $z=4.9$ with a similar Ly$\alpha$ EW as ID53.

\begin{figure}
\centering
        \includegraphics[width=9.4cm]{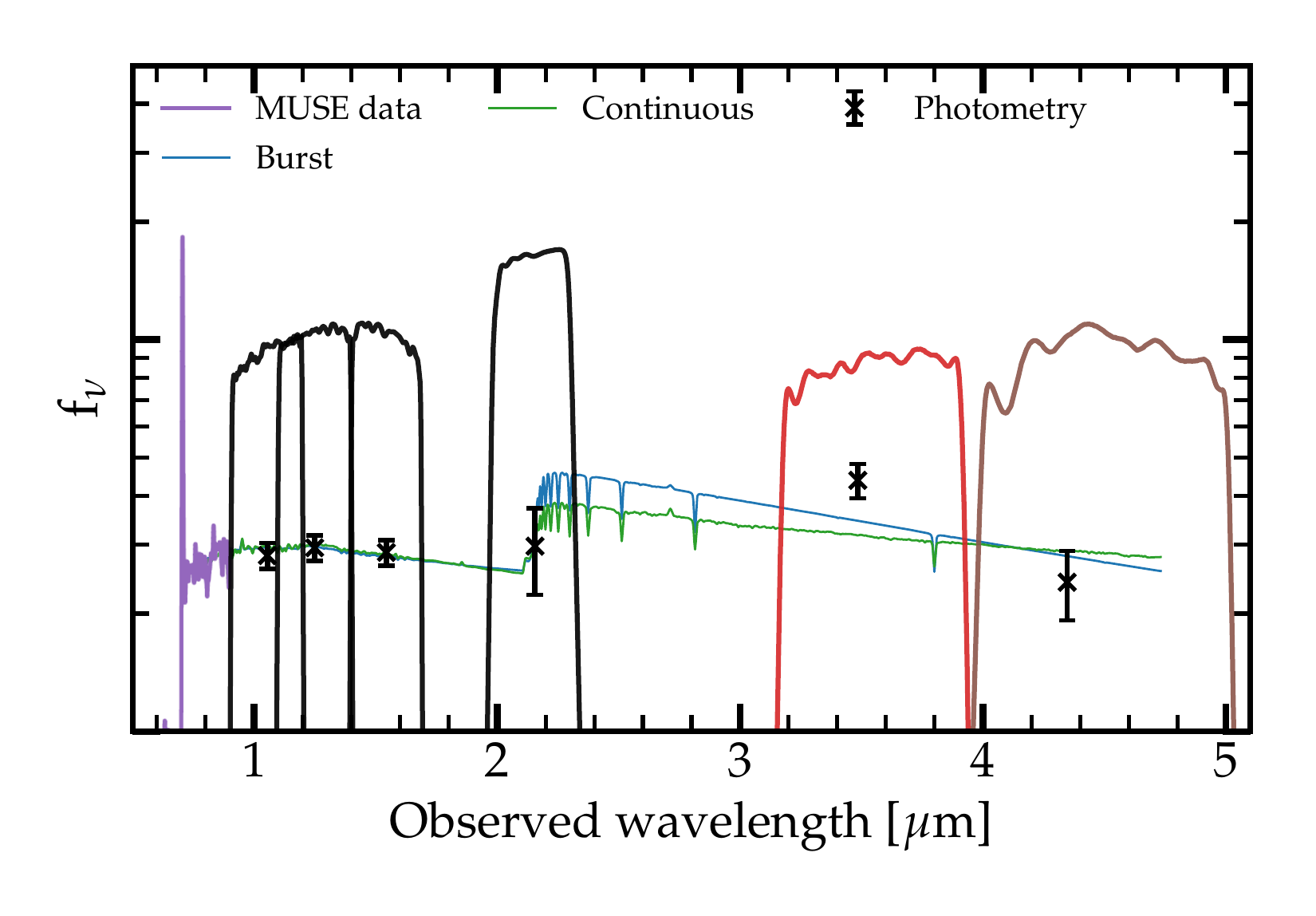} 
    \caption{SED of ID53 showing the MUSE data (purple line) and the black crosses show the photometry in the {\it HST}/WFC3 F105W, F125W, F160W, $K_s$ (black lines), and {\it Spitzer}/IRAC [3.6] and [4.5] filters (red and brown lines, respectively). Coloured lines show the best-fit SED models for the different SFHs. These models only include stellar light and no nebular line or continuum emission and therefore predict a lower limit to the observed {\it Spitzer} fluxes. We highlight the excess in the [3.6] filter compared to the [4.5] filter, which is used to estimate the H$\alpha$ line luminosity. }
    \label{fig:SED}
\end{figure}

\section{Synthesis: Physical picture of the galaxy ID53} \label{sec:synthesis}
We synthesise the measurements and analyses presented in Sections $\ref{sec:what}$, $\ref{sec:stellar_model}$ and $\ref{sec:nebular}$ in order to sketch a physical picture of the galaxy ID53.
ID53 is the brightest observed galaxy at $z>3$ in the coverage of the MXDF, but due to the limited cosmic volume probed ($\approx10^4$ cMpc$^3$), this corresponds to a fairly low-mass galaxy (M$_{\rm star}\approx10^9$ M$_{\odot}$). The expected number of ID53-like galaxies (i.e. within $\pm0.1$ magnitude difference) in an average region of the Universe with the volume of the MXDF at $z=4.8\pm0.5$ is 0.2 according to its UV luminosity \citep{Bouwens2021}, while it is only 0.01 according to its Ly$\alpha$ luminosity \citep{Herenz2019}. This suggests that ID53 is undergoing a relatively rare star burst, boosting its Ly$\alpha$ line luminosity with respect to the UV continuum. 

Our modelling of the UV spectrum suggests that the light is dominated by either a young star-burst or by a short bursty period that lasted for $<100$ Myr. The presence of particularly young (a few Myr) stars is indicated by the nebular C{\sc iv} emission and the stellar N{\sc v} P-Cygni feature. The star burst is plausibly driven by a recent merger event, as indicated from the {\it HST}/ACS morphology, the physical separation of $\approx1$ kpc, and the $\approx50$ km s$^{-1}$ velocity offset between the two components \citep[e.g.][]{Zanella2021}.

The age of the stars that dominate the UV light is only $\approx20-40$ Myr, suggesting that it formed its stars with a typical star formation rate (SFR) of $\approx100$ M$_{\odot}$ yr$^{-1}$, well above the typical SFR for galaxies with such masses at $z\approx5$ \citep[e.g.][]{Stark2013,Santini2017}. When standard calibrations between the SFR and the UV luminosity are applied, the unobscured SFR is 10 M$_{\odot}$ yr$^{-1}$ (see \S $\ref{sec:phot}$), which places the total SFR at least a factor 3 above the typical SFR for galaxies with a mass of $\approx10^9$ M$_{\odot}$ at $z\approx5$. Following \cite{Schreiber2015}, the fraction of galaxies that is expected to be in this star-bursting regime is $\lesssim5$ \%, which is in agreement with the rare nature of the star burst as inferred from the relative Ly$\alpha$ and UV number densities. A contribution from dust-obscured star formation would only increase the rarity of the star burst, but we note that galaxies with a mass comparable to that of ID53 typically show little obscured star formation \citep[e.g.][]{Whitaker2017}. A significant amount of dust would also be at odds with the strong Ly$\alpha$ line \citep[e.g.][]{Atek2008,Matthee2021}.

\begin{figure}
\centering
        \includegraphics[width=9.3cm]{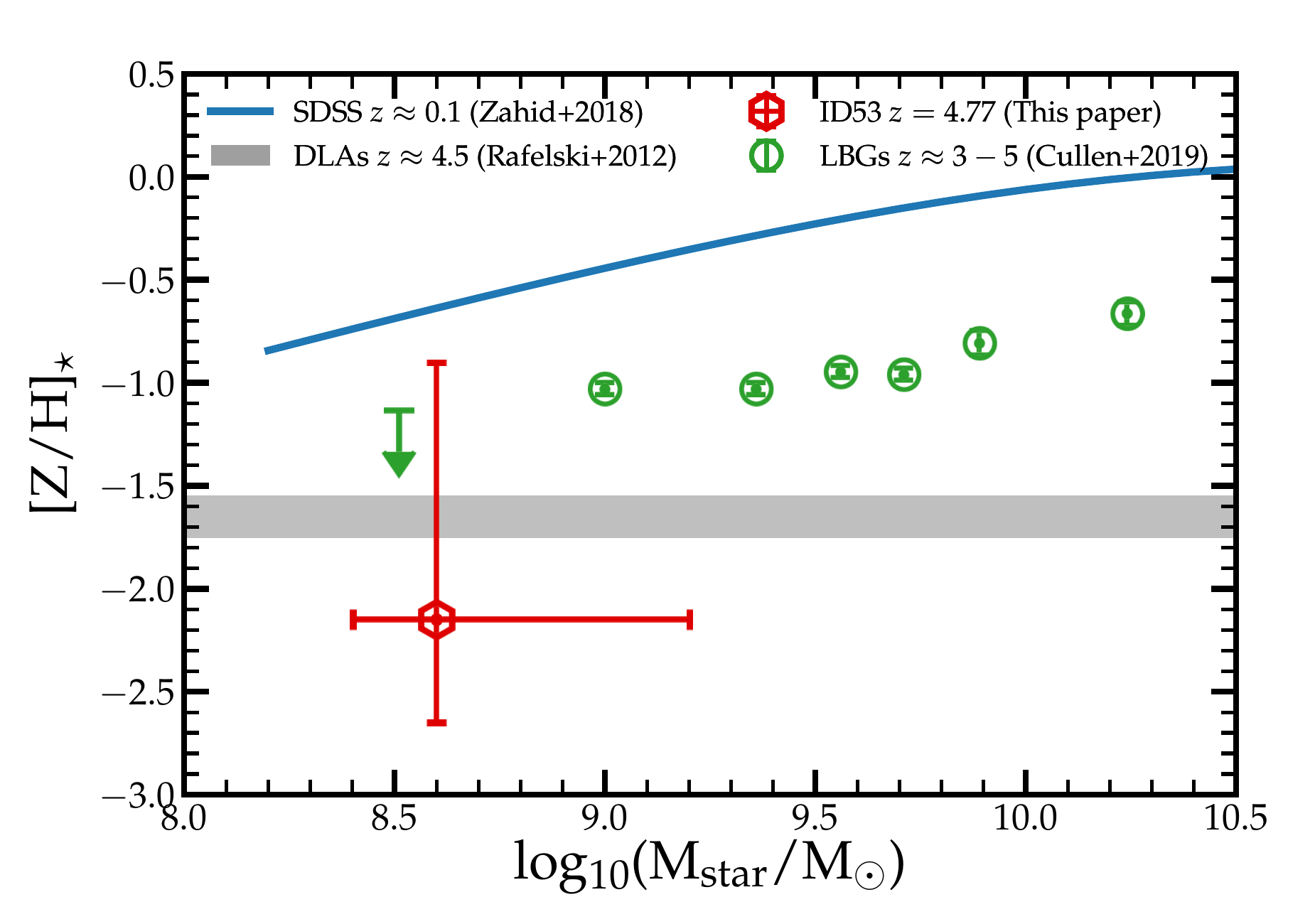} 
    \caption{Stellar mass-stellar metallicity relation over $z\approx0.1$ \citep{Zahid2017} to $z\approx3-5$ \citep{Cullen2019}. The green data points show measurements by \citet{Cullen2019} shifted down by 0.1 dex to account for differences in the stellar population models. The grey shaded region shows the average gas-phase metallicity measured in DLAs at $z\approx4.5$ \citep{Rafelski2012}. }
    \label{fig:ZstarMstar}
\end{figure}

The metallicity of the galaxy is [Z/H]$=-2.15^{+1.25}_{-0.5}$, which despeite the uncertainties suggests that it is well below the typical metallicity measured for more massive galaxies at $z=3-5$ (Fig. $\ref{fig:ZstarMstar}$). This suggests that the stellar mass-metallicity relation either has significant scatter, or that the mass dependence is much stronger at low masses than it is at higher masses. It is plausible that the relatively low metallicity of ID53 is connected to the young age of the galaxies. In such young galaxies, earlier generations of stars had less time to enrich the accreted gas. More accurate measurements of the metallicity are required for a more detailed investigation.

\section{Why the observed Ly$\alpha$ EW traces stellar metallicity} \label{discuss:EW}
It has been proposed that galaxies with very strong Ly$\alpha$ emission could be the hosts of (near)-primordial systems, \citep[e.g.][]{PartridgePeebles1967,Raiter2010}. The intrinsic\textup{} Ly$\alpha$ EW produced by a stellar population is very sensitive to age, with a secondary dependence on metallicity. In particular, galaxies with an intrinsic Ly$\alpha$ EW above $\gtrsim200$ {\AA} require extremely young ages and low metallicities ($\sim10^6$ yr and Z$\lesssim 0.1$ Z$_{\odot}$; see e.g. Figure 12 in \citealt{Hashimoto2017}). Galaxies with a high obseved Ly$\alpha$ EW therefore have attracted significant attention \citep[e.g.][]{MalhotraRhoads2002,Kashikawa2012,Sobral2015,Maseda2020}. Because of resonant scattering, the attenuation of Ly$\alpha$ photons is often higher than the attenuation of the UV continuum, lowering the observed EW \citep[e.g.][]{Scarlata2009,Henry2015,Yang2017}. On the other hand, other possible sources of Ly$\alpha$ emission exist, such as collisional ionisation \citep{Dijkstra2006} and fluorescent recombination emission from external sources \citep{Cantalupo2012,Marino2018}. ID53, however, is not in the vicinity of a bright quasar, and the narrowness of its Ly$\alpha$ line-profile is at odds with a significant contribution from collisional ionisation. The observed EW of 62 {\AA} of ID53 \citep{Leclercq2017} can therefore realistically be considered as a lower limit of the intrinsic EW.

\begin{figure}
\centering
        \includegraphics[width=9.3cm]{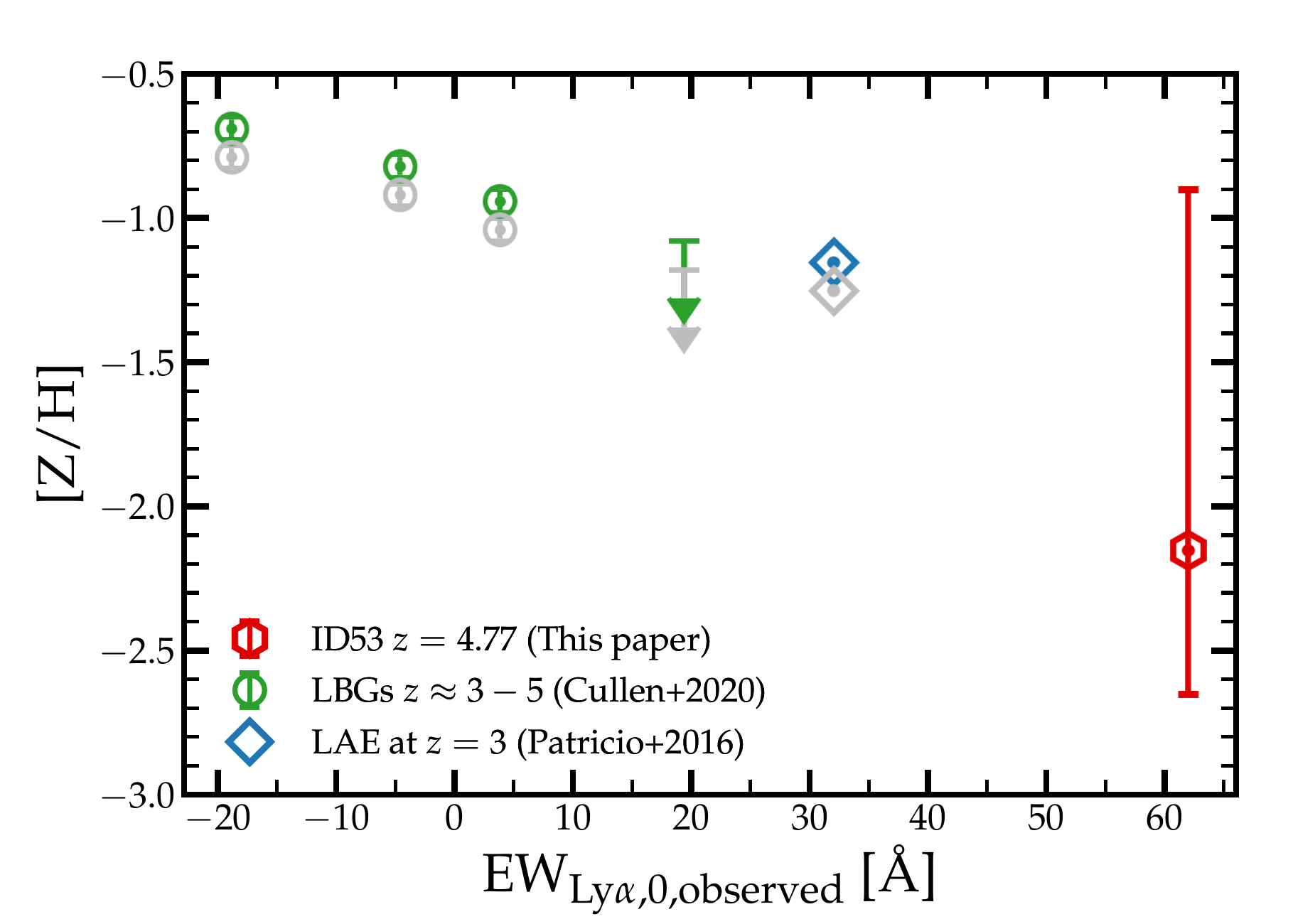} 
    \caption{Observed rest-frame Ly$\alpha$ EW vs. the stellar metallicity inferred from the UV spectrum. For ID53 we show the light-weighted metallicity preferred by the models with a complex SFH. The metallicities inferred for the LAEs from \citet{Patricio2016} and this study show that the trend between stellar metallicity and Ly$\alpha$ EW identified by \citet{Cullen2020} can be extrapolated to higher EWs. The grey data points show measurements by \citet{Patricio2016} and \citet{Cullen2020} shifted by 0.1 dex to account for differences in the assumed stellar population models. }
    \label{fig:EW_comp}
\end{figure}

We compare our inferred stellar metallicity estimate of ID53 in relation to its Ly$\alpha$ EW to other recent results in Fig. $\ref{fig:EW_comp}$. We compiled metallicity estimates for UV-selected galaxies binned by their Ly$\alpha$ EW \citep{Cullen2020} and the results from \cite{Patricio2016} of a lensed bright Ly$\alpha$ emitter at $z=3.5$.  \cite{Cullen2020} assumed a continuous SFH with a fixed age of 100 Myr (in which case we would obtain [Z/H]$-2.15^{+0.0}_{-1.0}$). The stellar metallicities in \cite{Patricio2016} and \cite{Cullen2019} were measured with Starburst99 models, which result in metallicities that are systematically 0.1 dex higher \citep{Cullen2019}. Fig. $\ref{fig:EW_comp}$ shows that the correlation between observed EW and metallicity as noted by \cite{Cullen2020} can be extrapolated towards higher EWs. This anti-correlation may partly follow trends between Ly$\alpha$ EW and mass \citep[e.g.][]{Oyarzun2017}, but as discussed in \S $\ref{sec:synthesis},$ the galaxy has a relatively low metallicity for its mass.

The trend between observed Ly$\alpha$ EW and metallicity therefore implies that the observed EW traces the chemical maturity of a galaxy as it is sensitive to various processes: i) a less mature galaxy has a higher intrinsic EW as the stars are younger and more metal poor, and ii) a less mature galaxy has less dust and therefore a higher Ly$\alpha$ escape fraction \citep[e.g.][]{Atek2008}, leading to a higher EW \citep{SM2019}. \cite{Calabro2021} showed that galaxies with bluer UV slopes (i.e. those with a lower attenuation) tend to have a lower stellar metallicity (and therefore are indeed less mature). Additionally, it may be possible that less mature galaxies may also have a lower HI column density, which affects the relative escape of Ly$\alpha$ photons compared to UV continuum photons \citep[e.g.][]{Yang2017}. This empirical result therefore confirms that galaxies with very high observed Ly$\alpha$ EWs \citep[e.g.][]{Maseda2020} are promising targets that could trace the most metal-poor and primordial star bursts.

\section{Future directions and caveats} \label{sec:discuss}
We have shown that single-burst and models with continuous star formation histories are almost equally well capable of fitting the UV spectrum of ID53. In \S $\ref{sec:Halpha},$ we showed that their H$\alpha$ luminosities offer great opportunity to differentiate these models (see also Appendix $\ref{sec:ApC}$), in particular when the attenuation is known. Here we discuss the other observables that might offer additional constraining power,  more complicated model additions, and variations, and which caveats underly our work.

\begin{figure}
\centering
        \includegraphics[width=9.3cm]{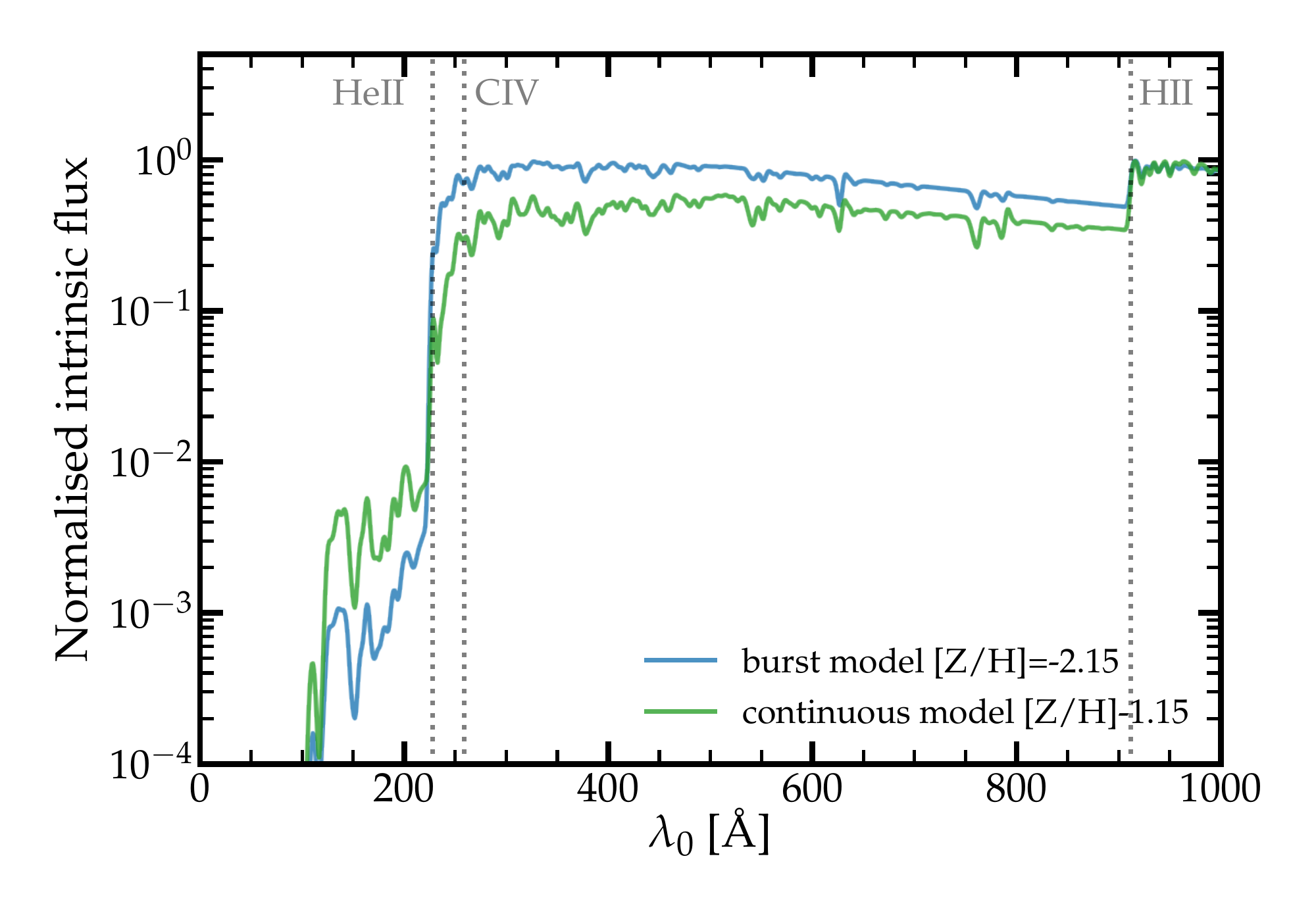} 
    \caption{Extreme intrinsic{\it } UV spectrum of the best-fitting single-burst (blue) and continuous SFH (green). These are the spectra of the best-fit models without dust attenuation and normalised to the continuum level at $\lambda_0=912 {\AA}$. Dotted grey lines mark the wavelength corresponding to the ionisation energies of H{\sc ii}, C{\sc iv} and He{\sc ii} (from right to left, respectively).}
    \label{fig:relative_HeII_CIV}
\end{figure}

\subsection{Additional emission lines}
It is expected that differences in stellar populations impact various nebular emission-line ratios \citep[e.g.][]{Strom2018,Topping2020}. As we assumed that attenuation within H{\sc ii} regions is negligible, we compared the ionising continuum of the intrinsic SEDs of various models. In Fig. $\ref{fig:relative_HeII_CIV}$ we show these continua normalised to the non-ionising continuum level at 912 {\AA}. The hardness of the spectrum differs between the models, which impacts the strengths of various high-ionisation emission lines such as C{\sc iv} and He{\sc ii} \citep[e.g.][]{Feltre2016,Berg2021}. Fig. $\ref{fig:relative_HeII_CIV}$ shows that there are significant differences at wavelengths for which the photon energies (i.e. $>4$ Ry) are high enough to ionise helium. The stellar+nebular He{\sc ii}$_{1640}$ feature, unfortunately just outside the MUSE wavelength coverage for ID53, will be the strongest in case of continuous star formation. Remarkably, the two-burst model, which has the lowest flux density in the $\lambda=300-900$ {\AA} range, has the second highest flux density at $>4$ Ry, and vice versa for the single burst.

A detailed inclusion of the measured C{\sc iv} EW in order to differentiate stellar models is beyond the scope of this paper as their predicted C{\sc iv} luminosity is also affected by the metallicity, the C/O abundance, and the gas density \citep[e.g.][]{Nakajima2018}. The latter issues are less important for He{\sc ii}, but we note that it is known that several galaxies are observed to have a nebular He{\sc ii} EW that is too high to explain it with these types of models \citep[e.g.][]{Plat2019,StanwayEldrige2019}. Nevertheless, emission line ratios that trace the relative abundance of photons with energies in the 2-4 Ry range (e.g. C{\sc iv}/He{\sc ii} and He{\sc ii}/H$\alpha$) offer significant constraining power for models that similarly well describe the non-ionising part of the UV spectrum, in particular when the density and metallicity of the gas are known from detailed spectroscopy of a suite of emission lines. 

\subsection{Complex star formation histories} \label{sec:discuss_complex}
As shown in \S $\ref{sec:phot}$, two distinct regions are visible within ID53 with relative luminosity ratios of $1:2$ in the {\it HST} data in the F814W filter (Fig. $\ref{fig:HST}$). As shown by the red contours in Fig. $\ref{fig:HST}$, the continuum is significantly blended in the MUSE data, so that we cannot separately fit the UV spectra of the different components. Nevertheless, it is interesting to consider the possibility that the two components seen in the {\it HST} data correspond to stellar populations that are non-coeval. We find that it is not possible to identify statistically significant differences between the ages and metallicities of the stars in these components. This is likely because the 1D spectra extracted centred on these components are significantly blended.

Qualitatively, differences in the ages and/or metallicities of the components are plausible because the nebular C{\sc iv} emission line is much stronger in the fainter component (\S $\ref{sec:CIVlines}$). This suggests that the fainter component is younger (see the previous subsection and references therein). It is possible to explain the observed spectrum of ID53 with a stellar spectrum composed of a complex SFH with more flexibility than the single-burst or continuous star formation. Such more complex models could be models with two bursts, or a continuous star formation followed by a burst. However, the main challenge of explaining the observed spectrum by invoking more complex star formation histories is that they introduce more than twice the number of free parameters. Since the $\chi_{\rm red}^2$ of the two different models with simple star formation histories are already very similar, it is therefore not warranted to perform full spectral fitting with complex star formation histories without making a range of strong assumptions. 

\begin{figure}
\centering
        \includegraphics[width=8.8cm]{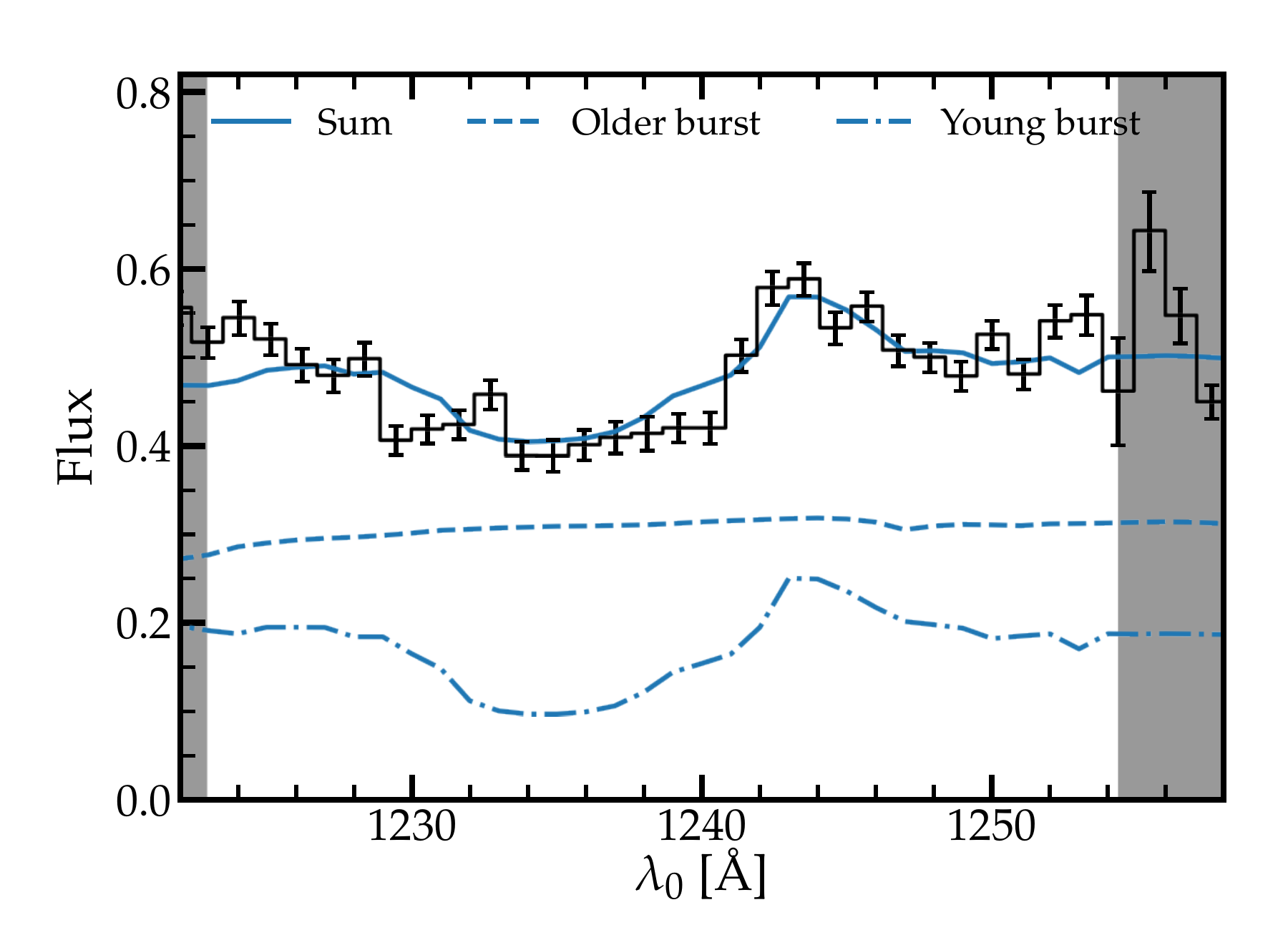} 
    \caption{Zoom-in on the N{\sc v} P-Cygni profile of ID53. Here we show the combined and individual model spectra of an example model with a two-component SFH. We stress that this model is not a unique solution and acts as an example for the discussion in \S $\ref{sec:discuss_complex}$. This model could be applicable to ID53 because of its clumpy appearance in the {\it HST} morphology and could solve the possible difference between the intrinsic and inferred H$\alpha$ luminosities (\S $\ref{sec:Halpha}$).}
    \label{fig:2comp_example}
\end{figure}

A full detailed analysis of complex star formation histories in ID53 requires the inclusion of nebular emission lines and/or spatially resolved information. Nevertheless, as an informative and notably fine-tuned example, we show in Fig. $\ref{fig:2comp_example}$ how the N{\sc v} P-Cygni profile of ID53 can be matched by a model that consists of two different populations: a relatively faint young component that shows the N{\sc v} profile, and a brighter older component that shows no stellar wind features. The relative flux ratio of these two components is 1:2, similar to the flux ratio seen in the {\it HST} image, and the fainter component greatly dominates the ionising luminosity above 3 Ry (i.e. the energies required to emit nebular C{\sc iv}). It is particularly interesting to mention that the combined H$\alpha$ luminosity of this model is $\approx10^{42.7}$ erg s$^{-1}$, that is, a factor $\approx7$ lower than the H$\alpha$ luminosity from the models with a single-burst and continuous star formation (Table $\ref{tab:chisel}$) and much closer to the observed H$\alpha$ luminosity (see \S $\ref{sec:Halpha}$). The reason for the lowered H$\alpha$ luminosity is that the older population contributes significantly to the UV light, but does not contribute significantly to the ionising luminosity that controls the H$\alpha$ output. 
While we think this is a plausible scenario providing a good description of the MUSE data, the H$\alpha$ luminosity, and the apparent morphology, we stress that this model is definitely not the only unique model that can do so, and it will be interesting to distinguish between this and other models in future work when more data are available.

\subsection{Caveats}
The first main caveat underlying our work is that our results are dependent on the validity of the BPASS stellar population models in the UV part of the spectrum. Other stellar population models, such as Starburst99 \citep{Leitherer1999}, have different spectral shapes around (e.g.) the N{\sc v} P-Cygni feature; see for example Fig. 13 in \citealt{StanwayEldrige2019}. The terminal velocities and masses of stellar outflows that impact P-Cygni profiles are particularly important for stellar wind features \citep[e.g.][]{Vink2001}. The models are uncertain because of theoretical uncertainties in the evolution of massive stars in binaries \citep[e.g.][]{Gotberg2019}, rotation velocities of massive stars \citep[e.g.][]{Levesque2012}, and observational challenges in obtaining UV spectroscopy of individual and binary massive stars \citep[e.g.][]{Crowther2016,Holgado2020}. On the other hand, these BPASS models have proven to result in the best fits for typical galaxies at $z\approx2-3$ \citep{Steidel2016}. BPASS models without the inclusion of binary stars also yield a somewhat poorer match to the MUSE spectrum of ID53 (Appendix $\ref{sec:IMFvar}$). In order to address this caveat in the future, new detailed observations of very distant galaxies will need to be accompanied by observations of local analogous galaxies \citep[e.g.][]{Kehrig2015,Senchyna2017,Senchyna2021,Berg2019,Wofford2021} and H{\sc ii} regions in the (very) nearby Universe \citep[e.g.][]{Castro2018} that can allow the detailed resolved observations required to constrain models describing the properties of massive stars at very low metallicities in detail.

A second caveat is the shape and upper mass limit of the IMF. We have assumed a standard Chabrier IMF with an upper mass limit of 100 M$_{\odot}$ and note that this results in good fits to the integrated UV spectrum. In the metallicity range [Z/H]$\lesssim-2,$ it is theoretically expected that the IMF is top heavy compared to the Chabrier IMF \citep[e.g.][]{Chon2021}. As detailed in \cite{StanwayEldrige2019}, the number of ionising photons can vary by an order of magnitude when changing the upper mass or the slope of the IMF. Except for the youngest ages when extremely high-mass stars could be present, the slope in particular is important. In Appendix $\ref{sec:IMFvar}$ we present fits to the spectrum using BPASS models with various IMF variations. Our modelling cannot distinguish between the fiducial IMF and, for example, a Chabrier IMF with a higher mass cut-off of 300 M$_{\odot}$ as the $\chi^2$ are very similar. Variations in the IMF are to some extent degenerate with variations in the SFH and mostly impact the age, stellar mass, and H$\alpha$ luminosity (see Appendix $\ref{sec:IMFvar}$); they have a smaller impact on the metallicity. The current data do not offer sufficiently detailed sensitivity to the number of hydrogen and helium ionising photons that is required to investigate variations in IMF at the very massive end, while simultaneously also allowing for variations in the SFH. We therefore defer such a detailed investigation of ID53 to future work when more data (e.g. rest-frame optical spectroscopy and further UV spectroscopy in the $\lambda_0=1600-3000$ {\AA} range) are available. Furthermore, we remark that stochastic IMF sampling \citep[e.g][]{VidalGarcia2017} should have a minor role as the star bursts in our analysis have masses $>10^7$ M$_{\odot}$.

A final caveat to discuss is the possible contribution from an AGN to the rest-frame UV continuum and line emission. Narrow-line or type II AGN typically show a relatively red spectrum with a UV slope $\beta\approx-0.3$ \citep{Hainline2011} apart from Ly$\alpha$, N{\sc v}, Si{\sc iv}, N{\sc iv}] and C{\sc iv} emission in the wavelength range covered by the MUSE data, with Ly$\alpha$ and C{\sc iv} being the strongest similar to ID53. Some relatively blue ($\beta\sim-2$) Type II AGN with similar emission lines have recently been discovered \citep[e.g.][]{Lin2021,Tang2021}. Similar to the multiple stellar components discussed in \S $\ref{sec:discuss_complex}$,  one of the components might be an AGN. AGN continuum emission would be an additional component to the total spectrum and therefore decrease the strength of stellar absorption and emission features in the combined or blended spectrum. This could impact the stellar population modelling results. However, we argue that it is unlikely that this is the case for ID53 because the Ly$\alpha$ and C{\sc iv} emission lines are considerably narrower (by a factor $>5$) than the line widths measured in blue narrow-line AGN \citep[$\approx500$ km s$^{-1}$, e.g.][and references above]{Sobral2018}. Furthermore, the C{\sc iv} emission spatially coincides with the fainter component that appears to be marginally resolved in further substructure (Fig. $\ref{fig:HST}$), instead of coinciding with the relatively point-like brighter component, which would be expected for such an AGN. Future spatially resolved rest-frame optical emission-line diagnostics (i.e. the BPT diagram) could further consolidate this.

\section{Summary} \label{sec:summary}
We studied the galaxy ID53 at $z=4.7745$, the UV-brightest star-forming galaxy at $z>3$ in the MUSE eXtremely Deep Field. The galaxy was observed for a total of 140 hours. ID53 has a typical L$^{\star}$ UV luminosity with a relatively blue continuum slope and a strong Ly$\alpha$ line with EW=62 {\AA}. The rest-frame UV morphology as seen by {\it HST}/ACS reveals a clumpy structure that is mostly unresolved in the MUSE data, which trace $\lambda_0=900-1600$ {\AA} (Fig. $\ref{fig:1dspec}$). Our results are listed below.

\begin{itemize}

\item We measured the systemic redshift of ID53 with the faint non-resonant fine-structure lines O{\sc i}*, Si{\sc ii}* , and C{\sc ii}* (Fig. $\ref{fig:emission}$ and Table $\ref{tab:emlines}$). The measured systemic redshift implies that the peak of the Ly$\alpha$ line is redshifted with respect to the systemic redshift by $+195$ km s$^{-1}$, which is a typical value for galaxies with comparable Ly$\alpha$ EW. 

\item The nebular C{\sc iv} emission-line doublet is clearly detected with an interstellar absorption-corrected integrated EW of 4.0 {\AA} (Fig. $\ref{fig:carbon_absorption}$). The lines are narrow and redshifted by +51 km s$^{-1}$ with respect to the systemic. This redshift arises because C{\sc iv} particularly originates from the fainter north-western component of the galaxy (Fig. $\ref{fig:emission}$). This fainter component likely harbours the youngest stars. 

\item The stellar continuum of ID53 is detected with an S/N ranging from 13 to 33 per $\lambda_0=1$ {\AA} interval, with the higher sensitivity in the bluer part of the spectrum, where we detect a clear N{\sc v} P-Cygni feature (Fig. $\ref{fig:Pcyg}$). We modelled the stellar continuum by combining BPASS models for single stellar populations with either a burst or a continuous SFH and an attenuation curve with varying E$(B-V)$. We fitted these models to the rest-frame UV continuum spectrum and masked wavelengths with non-stellar features. IR photometry from {\it HST} and {\it Spitzer} was used as an additional bound (see \S $\ref{sec:stellar_model}$ for details). We showed that detailed features in the rest-frame UV continuum spectrum of ID53 can be well matched by BPASS stellar population models combined with a standard IMF, but the galaxy needs to be allowed to be younger than the 100 Myr that are typically assumed in this modelling. The best-fit metallicity and age are generally low and dependent on the assumed SFH (Table $\ref{tab:chisel}$). They range from [Z/H]=-2.15 to -1.15 and log$_{10}$(age/yr)=6.5-7.6. 

\item The UV continuum spectrum alone cannot sufficiently distinguish the SFH that best fits ID53. By combining likelihood distributions of various SFHs, we derived fiducial measurements of log$_{10}$(M$_{\rm star}$/M$_{\odot}$) = 8.6$^{+0.6}_{-0.2}$ and [Z/H]=$-2.15^{+1.25}_{-0.50}$. While uncertain, the stellar metallicity of ID53 appears to be lower than that of more massive galaxies at $z\sim5$. As the galaxy is undergoing a relatively rare star burst, this suggests that significant scatter exists in the mass - metallicity relation, where younger galaxies are in an earlier phase of their chemical enrichment (\S $\ref{sec:synthesis}$).

\item The H$\alpha$ luminosity that is inferred from {\it Spitzer}/IRAC photometry (\S $\ref{sec:Halpha}$) is almost an order of magnitude lower than the intrinsic H$\alpha$ luminosity predicted by the best-fit stellar population models, and it is lower than most of the fitted models with a continuous SFH. This could indicate a significant escape fraction of ionising photons, a high nebular attenuation compared to the stellar attenuation, or a more complex SFH (\S $\ref{sec:discuss}$).

\item We confirm empirical results that the stellar metallicity of high-redshift galaxies is anti-correlated with the observed Ly$\alpha$ EW at $z=3-5$ (\S $\ref{discuss:EW}$), extending the dynamic range of previous results. We argue that this trend is most likely driven by underlying correlations between the Ly$\alpha$ escape fraction and chemical maturity, for example because stellar metallicity may well correlate with the amount of dust. Galaxies with a high Ly$\alpha$ EW are therefore good targets to prioritise in observational searches for galaxies hosting extremely low metallicity stars.

\item We discussed that future measurements of the (spatially resolved) intrinsic H$\alpha$ luminosity and long-wavelength stellar continuum measurements will be very useful in further constraining the star formation histories (\S $\ref{sec:discuss}$). Furthermore, the lowest metallicity models currently allowed by the MUSE spectrum have significantly harder spectra, which impacts the strengths of high-ionisation lines as C{\sc iv} and He{\sc ii}. Detailed photo-ionisation modelling to include these data will be warranted if several parameters of the interstellar medium such as the density, temperature, and metallicity can be constrained from nebular emission-line measurements, and these will therefore be useful in future analyses of low-metallicity systems. 

\item Our work contains important caveats about the IMF and the validity of the UV spectra of the young and metal-poor stellar population models that we used. However, our results show that the current BPASS stellar population models can explain all key observed features of the stellar spectrum with a standard IMF. Furthermore, while the data cannot constrain various IMFs, the fitted metallicity is relatively insensitive to the adopted IMF (\S $\ref{sec:discuss}$). A contribution from a low-luminosity AGN is an additional caveat, but the narrowness of the emission lines and the spatially resolved emission-line morphology suggests that any such contribution is small.
\end{itemize}

Our results highlight the value of spatially resolved spectroscopy in obtaining a complete picture of typical galaxies in the early Universe. In the future, this study may be expanded to statistical samples of very high-redshift galaxies with sensitive spectroscopy using 30 m class telescopes. The success of detailed studies of their stellar population will depend on flexibility in star formation histories and the availability of data  and methods to constrain this. This means that galaxies need to be observed over a wide wavelength range (in particular including rest-frame optical emission lines). The lower the metallicity of a galaxy, the harder it is to constrain (because photospheric and stellar wind features are weaker). Therefore, the simultaneous inclusion of nebular emission line ratios through detailed photo-ionisation modelling will be particularly relevant in future studies of the first generations of stars.

\section*{Acknowledgements} 
We thank the referee for thoughtful and constructive comments that have improved the quality of this manuscript.
Based on observations collected at the European Southern Observatory under ESO programme 1101.A-0127.
This work made use of v2.2.1 of the Binary Population and Spectral Synthesis (BPASS) models as described in \cite{Eldrige2017} and \cite{BPASS2018}. AF acknowledges the support from grant PRIN MIUR2017-20173ML3WW\_001. TN acknowledge support from Australian Research Council Laureate Fellowship FL180100060.

\bibliographystyle{aa}

\bibliography{MXDF.bib}

\begin{appendix}

\FloatBarrier
\section{Best stellar population model fits as a function of metallicity}
For completeness and detailed comparison, we list the properties of the best-fit stellar population models for each stellar metallicity in Table $\ref{tab:chisel_complete}$. To emphasise how well each spectrum matches the N{\sc v} and C{\sc iv} P-Cygni features, we also list the $\chi^2_{\rm red}$ when these features were considered alone.

\begin{table}
\centering
\caption{SED fitting: Full results.} \label{tab:chisel_complete} 
\begin{tabular}{lrrrrr} \hline
[Z/H] & $\Delta \chi^2_{\rm red}$ & log(age) & $\chi^2_{\rm red, NV}$ & $\chi^2_{\rm red, CIV}$ & log(L$_{\rm H\alpha, int}$) \\
& & & & &  \\
Burst & & & & &  \\
-3.15 & +0.77 & 6.4 & 13.5 & 4.01 & 43.93     \\
-2.15 & 0.0 & 6.5 & 4.33 & 4.26 & 43.58  \\
-1.15 & +0.86 & 6.6 & 6.54 & 3.87 & 43.15   \\
-0.85 & +1.27 & 6.6 & 6.75 & 10.42 & 43.13   \\
-0.54  & +2.12 & 6.5 & 7.28 & 18.96 & 43.21  \\
-0.37  & +2.61 & 6.5 & 6.19  & 26.44 & 43.15  \\
-0.24  & +3.18 & 6.5 & 6.61 & 29.93 & 43.20  \\
-0.15 & +5.61 & 6.6 & 19.94 & 25.73 & 43.18  \\
0.00  & +5.94 & 6.5 & 21.60 & 24.30 &  43.03  \\ 
& & & & \\

Cont. & & & & &  \\
-3.15 & +0.45 & 7.4 & 11.57 & 5.81 & 43.64  \\
-2.15 & +0.03 & 6.9 & 5.31 & 4.09 & 43.64 \\
-1.15 & 0.0 & 7.3 & 3.52 & 4.67 & 43.30  \\
-0.85 & +0.34 & 7.3 & 3.55 & 7.14 & 43.26\\
-0.54 & +1.28 & 7.3 & 5.10 & 11.60 & 43.20  \\
-0.37 & +2.04 & 7.3 & 7.03 & 14.48 & 43.14  \\
-0.24 & +2.73 & 7.3 & 8.78 & 17.03 & 43.10   \\
-0.15 & +3.32 & 7.6 & 11.58 & 13.33 & 42.86   \\
0.00 & +4.03 & 7.3 & 14.87 & 16.49 & 43.08 \\
\hline
\multicolumn{6}{p{.47\textwidth}}{\footnotesize We list the best-fit model for each stellar metallicity and for a single-burst or continuous star formation that satisfies the additional bounds from the IR photometry (F105W$>25.2$ and [4.5]$>25.3$). The age is in yr. We list the goodness of fit using the full spectral data relative to the best model for a certain SFH ($\Delta \chi^2_{\rm red,full}$). The best-fit $\chi^2{\rm red, full}$ are 3.00 and 3.17 for the single-burst and continuous star formation models, respectively. We also list $\chi^2_{\rm red}$ when only using wavelength layers around the N{\sc v} and C{\sc iv} P-Cygni profiles. For each model we also list the intrinsic H$\alpha$ luminosity in erg s$^{-1}$.}
\end{tabular}

\end{table}

\section{IMF variations}\label{sec:IMFvar}
In order to investigate the impact of the specific IMF on the stellar population modelling, we re-performed the fitting procedure (\S $\ref{sec:stellar_model}$) using BPASS models with different IMFs. The Chabrier IMF with upper mass limit 100 M$_{\odot}$ is the fiducial IMF. We also used the BPASS model with single stars, that is, excluding binary stars with the fiducial IMF (abbreviated as {sin, Chab100}). The following IMF variations are included: { Chab300} model uses a Chabrier IMF with an upper mass limit of 300 M$_{\odot}$. The {shallow-100} and {shallow-300} models have a shallower high-mass IMF slope of $\alpha=-2.0$ (as opposed to $\alpha=-2.3$ for the Chabrier IMF) and upper mass limits of 100 M$_{\odot}$ and 300 M$_{\odot}$, respectively. The  {steep-100} and  {steep-300} models have a stepper high-mass slope ($\alpha=-2.7)$ and the same two upper mass limits. These IMF variations include binary stars. We include both single-burst and continuous star formation histories.

Table $\ref{tab:chisel_full}$ shows the best-fit model for each model variation. Models without binary stars yield poorer fits to the spectrum. In general, variations in the IMF only lead to small to moderate differences in the best-fit parameters and the $\chi^2_{\rm red}$. This is likely because the best-fit ages are $\gtrsim3$ Myr when the most massive stars (i.e. $>100$ M$_{\odot}$) no longer exist (for a burst) or only mildly impact the spectrum (for continuous star formation). For a single burst, the best-fit metallicity is always [Z/H]=-2.15, while the age varies by 0.2 dex from $10^{6.3-6.5}$ yr. The H$\alpha$ luminosity varies by 0.5 dex but has relatively large uncertainties. The stellar mass is more affected, in particular for the IMFs with a steeper slope that require a higher stellar mass in order to fit the spectrum. For a continuous SFH, most IMF variations lead to younger and more metal-poor fits ([Z/H]=-2.15 versus -1.15), with a SFH that only lasted $<10$ Myr and an H$\alpha$ luminosity that is 0.5 dex higher. These results indicate that the best-fit stellar metallicity of ID53 is only weakly sensitive to the IMF and that the H$\alpha$ luminosity has significant constraining power in determining which IMF is more appropriate.

\begin{figure*}
\centering
\includegraphics[width=16.3cm]{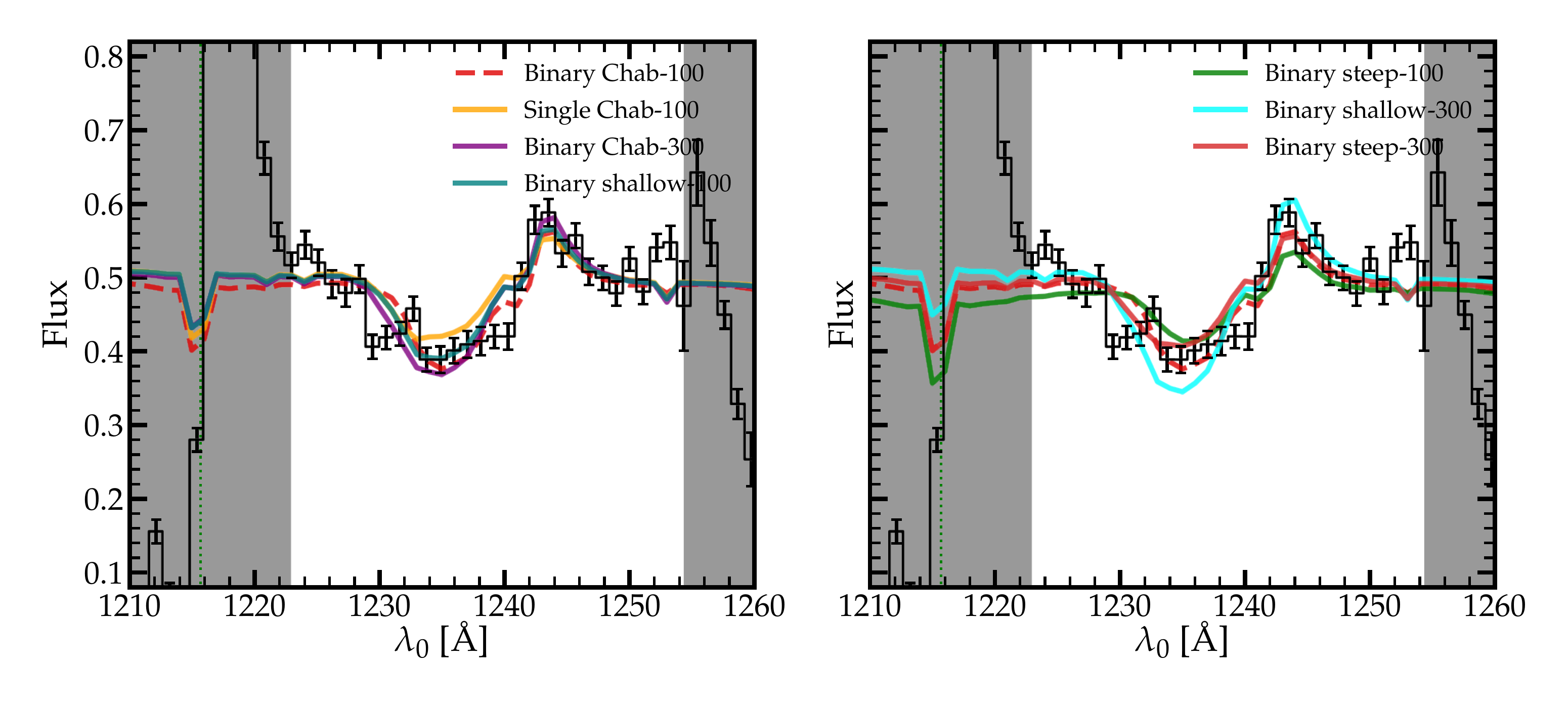} \\
\includegraphics[width=16.3cm]{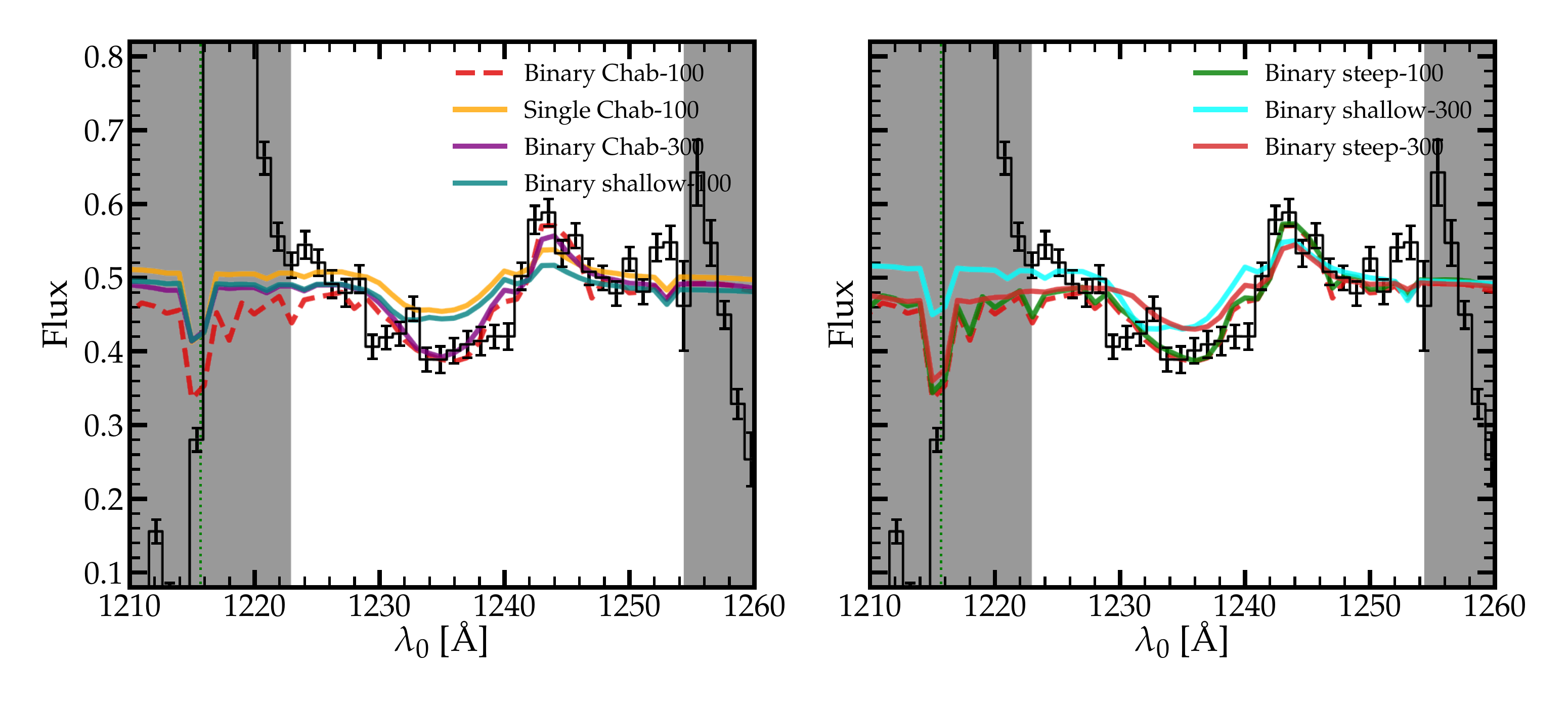} \\
\caption{Zoom-in on the N{\sc v} P-Cygni feature, where coloured lines show best-fit model spectra for various IMF variations and a single-burst (top) and continuous star formation (bottom). Data are shown in black, and grey shaded regions mark the regions ignored in spectral fitting due to nebular or interstellar absorption or emission. For a single burst, models match the N{\sc v} P-Cygni comparably well. For a continuous star formation, the model without binary stars and models with shallower IMFs yield poorer matches to the N{\sc v} profile. }
    \label{fig:IMF_PCygs}
\end{figure*}

For illustration, Fig. $\ref{fig:IMF_PCygs}$ shows the best-fit model for each IMF variation compared to the N{\sc v} P-Cygni feature observed in the spectrum. In general, single-burst models yield a similar match to the observed feature, while the models with single stars and a shallower IMF yield a poorer match to the N{\sc v} profile for the continuous SFH.

\begin{table*}
\centering
\caption{SED fitting: Model variations.}\label{tab:chisel_full}
\begin{tabular}{lrrrrrr} \hline

Type SFH  & [Z/H]$_{\rm full}$ & log$_{10}$(Age/yr) & log$_{10}$(M$_{\rm star}$/M$_{\odot}$) & log$_{10}$(L$_{\rm H\alpha, int}$/erg s$^{-1}$) & $\chi^2_{\rm red}$  \\ 

{\bf Burst, bin, Chab100}  & $-2.15^{+1.15}_{-1.0}$  & $6.5^{+0.6}_{-0.3}$ & $8.6^{+0.7}_{-0.2}$ & $43.6^{+0.3}_{-1.1}$ & 3.00   \\
Burst, sin, Chab100 & $-2.15^{+1.3}_{-1.0}$ & $6.4^{+1.1}_{-0.1}$ & $8.8^{+0.5}_{-0.4}$ & $43.5^{+0.6}_{-0.5}$ & 3.21 \\ 
Burst, bin, Chab300 & $-2.15^{+0.8}_{-1.0}$ & $6.3^{+0.7}_{-0.2}$ & $8.6^{+0.6}_{-0.4}$ & $43.8^{+0.3}_{-0.2}$ & 3.00 \\
Burst, bin, shallow-100 & $-2.15^{+1.15}_{-1.0}$ & $6.4^{+0.6}_{-0.4}$ & $8.5^{+0.7}_{-0.6}$ & $43.8^{+0.4}_{-0.4}$ & 3.03 \\ 
Burst, bin, shallow-300 & $-2.15^{+0.9}_{-1.0}$ & $6.3^{+1.0}_{-0.3}$ & $8.3^{+1.1}_{-0.6}$ & $43.3^{+0.7}_{-0.3}$ & 3.11 \\ 
Burst, bin, steep-100 & $-2.15^{+1.3}_{-1.0}$ & $6.5^{+0.7}_{-0.5}$ & $9.3^{+0.2}_{-0.4}$ & $43.4^{+0.4}_{-0.4}$ & 3.02\\  
Burst, bin, steep-300 & $-2.15^{+1.3}_{-1.0}$ & $6.3^{+0.8}_{-0.2}$ & $9.3^{+0.4}_{-0.4}$ & $43.8^{+0.1}_{-0.6}$ & 3.07 \\ 
\hline

{\bf Continuous, bin, Chab100 }  &  $-1.15^{+0.45}_{-2.0}$  & $7.3^{+0.6}_{-0.8}$ & $9.0^{+0.3}_{-0.4}$ & $43.5^{+0.4}_{-0.4}$ & 3.17   \\
Cont, sin, Chab100 & $-2.15^{+1.2}_{-1.0}$ & $6.5^{+0.2}_{-0.5}$ & $8.9^{+0.2}_{-0.2}$ & $43.9^{+0.2}_{-0.3}$ & 3.31\\ 
Cont,  bin, Chab300 & $-2.15^{+0.8}_{-1.0}$ & $6.5^{+0.2}_{-0.3}$ & $8.7^{+0.2}_{-0.2}$ & $43.9^{+0.1}_{-0.2}$ & 3.03 \\ 
Cont,  bin, shallow-100 & $-2.15^{+1.2}_{-1.0}$ & $6.4^{+0.1}_{-0.5}$ & $8.8^{+0.2}_{-0.3}$ & $44.1^{+0.1}_{-0.2}$ & 3.38 \\
Cont,  bin, shallow-300 & $-2.15^{+0.6}_{-1.0}$ & $6.3^{+0.1}_{-0.2}$ & $8.3^{+0.2}_{-0.1}$ & $43.9^{+0.2}_{-0.1}$ & 3.27 \\ 
Cont,  bin, steep-100 & $-1.15^{+0.4}_{-2.0}$ & $7.0^{+0.4}_{-1.0}$ & $9.3^{0.4}_{-0.1}$ & $43.3^{+0.7}_{-0.1}$ & 3.10 \\ 
Cont,  bin, steep-300 & $-2.15^{+1.4}_{-1.0}$ & $6.9^{+0.4}_{-0.9}$ & $9.3^{+0.2}_{-0.3}$ & $43.6^{+0.4}_{-0.3}$ & 3.08 \\ 

\hline
\multicolumn{6}{p{.87\textwidth}}{\footnotesize Results of the SED modelling of the rest-frame UV spectrum of ID53 for various different BPASS models and single-burst and continuous star formation histories. Columns as detailed in Table $\ref{tab:chisel}$. The fiducial results are listed in bold.}
\end{tabular}
\end{table*}

\section{Testing the influence of the S/N on stellar population modelling} \label{sec:ApB}
\begin{figure*}
\centering
        \includegraphics[width=16.3cm]{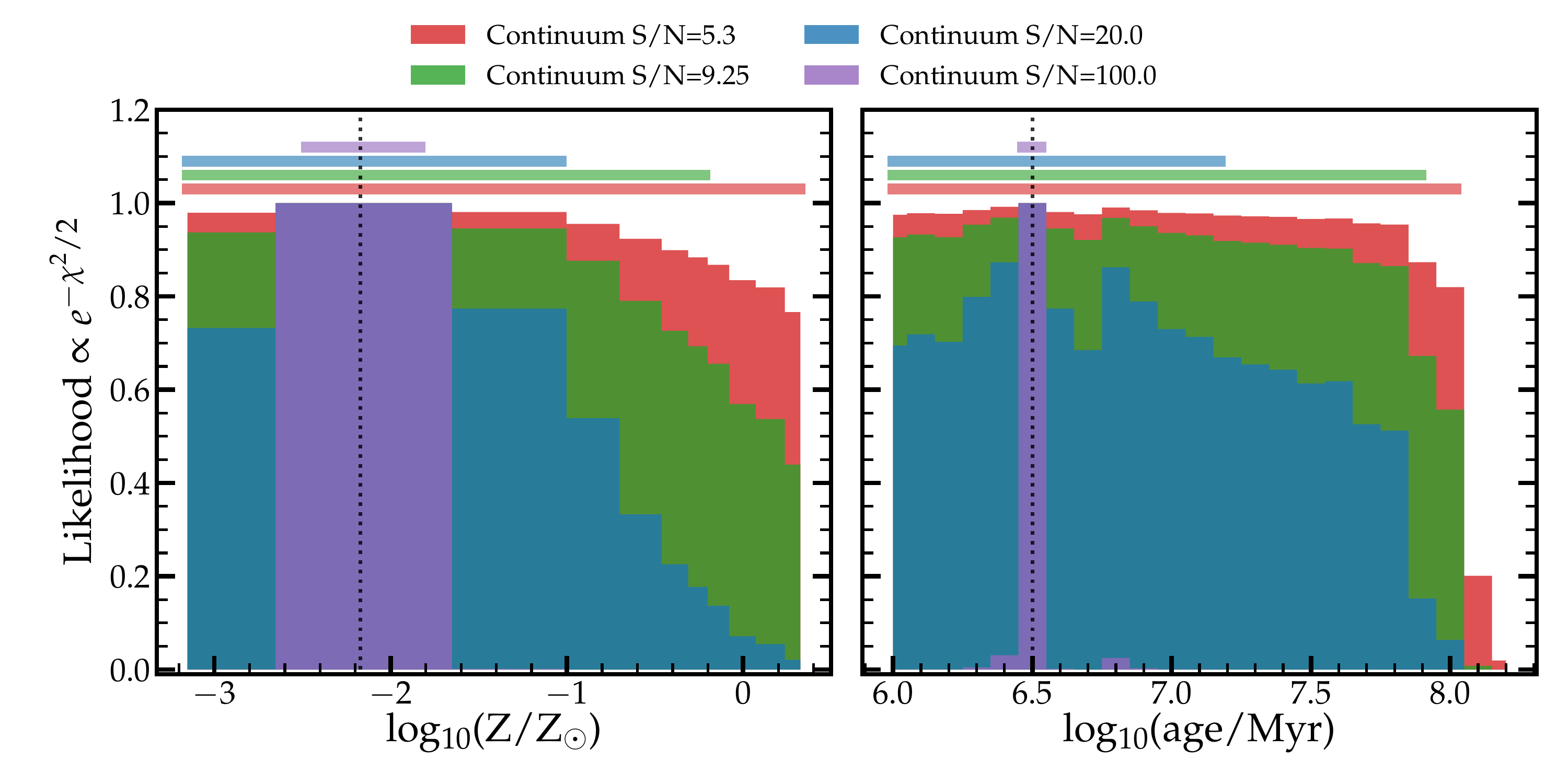}\\
    \caption{Impact of the sensitivity of the MUSE spectrum on our ability to constrain the age and metallicity of single-burst models of ID53. We simulated a single-burst model with [Z/H]=-2.15 and log$_{10}$(age/yr)=6.5 (illustrated with the dotted line) and derived the likelihood distribution with four different sensitivities. These correspond to exposure times of 10 hr (red), 30 hr (green), 140 hr (blue; i.e. these data), and 140 hr on a telescope that is five times larger (i.e. an MXDF on the ELT).  }
    \label{fig:model_SN}
\end{figure*}

Here we test the influence of the sensitivity of the MUSE spectrum on our ability to constrain the stellar metallicity and age of ID53.  For this simple experiment, we used the best-fit single-burst model with a metallicity of [Z/H]=-2.15 as a reference. We simulated spectra with a continuum S/N at the reference wavelength of $\lambda_0=1420$ {\AA} of 5.3, 9.25, 20, and 100. They correspond to a total exposure time of 10hr (the UDF-Mosaic), 30 hr (the UDF-10), 140 hr (i.e. these MXDF data), and an exposure time of 140 hr on a telescope with a mirror that is five times larger (i.e. an MXDF with MUSE mounted on the ELT) when we assumed for simplicity that the noise level scales with $t_{\rm exp}^{-1/2}$. Then, we calculated the likelihood for each model when compared to these simulated spectra, similarly as in the main text. We scaled the sensitivity of the MUSE data (including skylines) and only included the same wavelength layers as in the main text. No bounds on the photometry were used. In Fig. $\ref{fig:model_SN}$ we show the results. The MXDF data can rule out a metallicity of [Z/H]$>-0.85$ at $1\sigma$, while shallower data of the UDF-10 project (30 hour depth) would only have ruled out [ZH]$>-0.37$. Shallower data do not constrain the stellar metallicity at all. The right panel shows similar results for the stellar age. We also repeated this simple experiment for burst models with a higher metallicity. For these models, the variations in the spectrum are larger (see Fig. $\ref{fig:photospheric}$), and therefore shallower data can constrain these models with similar accuracy. 

\section{Testing the required H$\alpha$ S/N to constrain stellar population models} \label{sec:ApC}
\begin{figure*}
\centering
                \includegraphics[width=16.3cm]{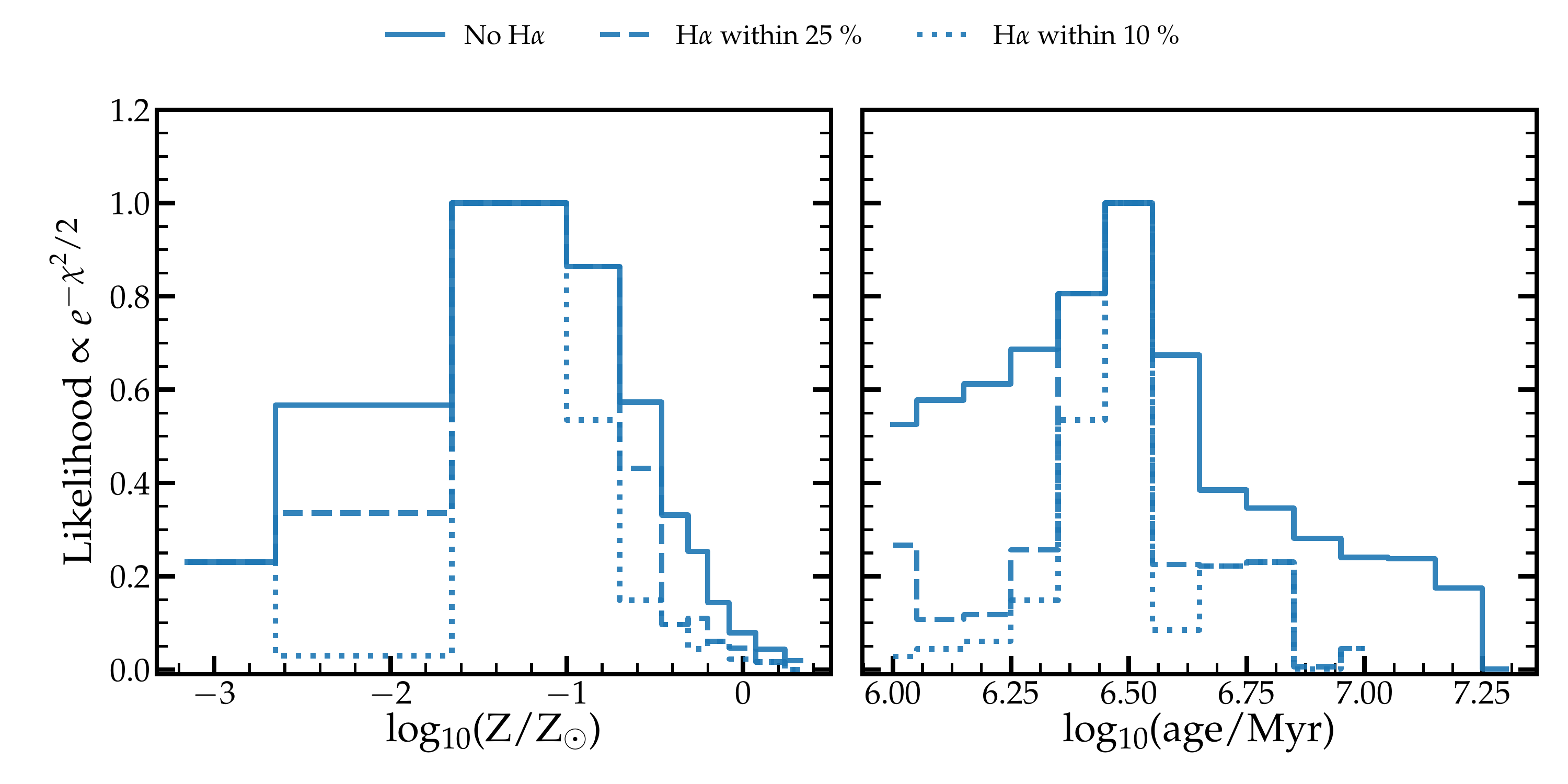}
    \caption{Impact of the availability and precision of the measurement of the intrinsic H$\alpha$ luminosity on our ability to constrain the age and metallicity of single-burst models of ID53. We simulated a single burst model with [Z/H]=-1.15 and log$_{10}$(age/yr)=6.5 and derived the likelihood distribution allowing all H$\alpha$ luminosities, or only those models that are within 25 and 10 \% of the real H$\alpha$ luminosity of the model.}
    \label{fig:model_Halpha}
\end{figure*}

Here we test how well we can improve our constraints on stellar population models if we have the additional information on the intrinsic H$\alpha$ luminosity produced by these stellar populations. We modelled a single-burst galaxy with [Z/H]=-1.15 and log$_{10}$(age/yr)=6.5 and fitted the stellar population models similar to the main text and Appendix $\ref{sec:ApB}$. Then, we added the additional bounds of only allowing models for the fitting with an H$\alpha$ luminosity within 25 and 10 \% of the real H$\alpha$ luminosity of the model. These different values were chosen to illustrate the effect of different S/N in the H$\alpha$ spectroscopy (including the dust and aperture corrections) and uncertainties in the conversion of the stellar models to the intrinsic H$\alpha$ luminosity (including recombination coefficients, dust within H{\sc ii} regions, and the escape fraction of ionising photons). The results are shown in Fig. $\ref{fig:model_Halpha}$. It is clear that even a 25 \% uncertainty on the H$\alpha$ luminosity would already significantly improve the uncertainties on the stellar population modelling.

\end{appendix}

\end{document}